\documentclass[journal]{IEEEtran}
\usepackage[T1]{fontenc}
\usepackage{cite}
\usepackage{amsmath,amssymb,amsfonts}
\usepackage{algorithm}
\usepackage{algorithmic}
\usepackage{graphicx}
\usepackage{textcomp}
\usepackage{xcolor}
\usepackage{subfigure}
\usepackage{booktabs}
\usepackage{multirow}
\usepackage{diagbox}
\usepackage{url}
\usepackage{threeparttable}
\usepackage{array}
\usepackage{textcase}
\usepackage{rotating}
\usepackage{etoolbox}

\makeatletter
\def\UrlAlphabet{%
      \do\a\do\b\do\c\do\d\do\e\do\f\do\g\do\h\do\i\do\j%
      \do\k\do\l\do\m\do\n\do\o\do\p\do\q\do\r\do\s\do\t%
      \do\u\do\v\do\w\do\x\do\y\do\z\do\A\do\B\do\C\do\D%
      \do\E\do\F\do\G\do\H\do\I\do\J\do\K\do\L\do\M\do\N%
      \do\O\do\P\do\Q\do\R\do\S\do\T\do\U\do\V\do\W\do\X%
      \do\Y\do\Z}
\def\UrlDigits{\do\1\do\2\do\3\do\4\do\5\do\6\do\7\do\8\do\9\do\0}
\g@addto@macro{\UrlBreaks}{\UrlOrds}
\g@addto@macro{\UrlBreaks}{\UrlAlphabet}
\g@addto@macro{\UrlBreaks}{\UrlDigits}

\patchcmd{\@algocf@start}
  {-1.5em}
  {0pt}
  {}{}

\makeatother

\newcommand{\SNR}[2]{${\text{SNR} #1 \text{#2}\;\text{dB}}$}
\def\post{\textit{a posteriori }}

\graphicspath{{figures/}}

\def\re{\mathrm{Re}}
\def\im{\mathrm{Im}}
\def\nt{N_{\rm t}}
\def\nr{N_{\rm r}}
\def\np{N_{\rm p}}
\def\vector{\mathrm{vec}}

\begin{document}
\setlength{\textfloatsep}{5pt}  
\setlength{\floatsep}{5pt}
\ifdefined \GramaCheck
  \newcommand{\CheckRmv}[1]{}
  \newcommand{\figref}[1]{Figure 1}%
  \newcommand{\tabref}[1]{Table 1}%
  \newcommand{\secref}[1]{Section 1}
  \newcommand{\algref}[1]{Algorithm 1}
  \renewcommand{\eqref}[1]{Equation 1}
\else
  \newcommand{\CheckRmv}[1]{#1}
  \newcommand{\figref}[1]{Fig.~\ref{#1}}%
  \newcommand{\tabref}[1]{Table~\ref{#1}}%
  \newcommand{\secref}[1]{Sec.~\ref{#1}}
  \newcommand{\algref}[1]{Algorithm~\ref{#1}}
  \renewcommand{\eqref}[1]{(\ref{#1})}
\fi
\newtheorem{theorem}{Theorem}
\newtheorem{proposition}{Proposition}
\newtheorem{assumption}{Assumption}
\newtheorem{definition}{Definition}
\newtheorem{condition}{Condition}
\newtheorem{property}{Property}
\newtheorem{remark}{Remark}
\newtheorem{lemma}{Lemma}
\newtheorem{corollary}{Corollary}
%
\title{Generative Diffusion Models for \\ High Dimensional Channel Estimation}

%
%
\author{Xingyu~Zhou,~\IEEEmembership{Graduate Student Member,~IEEE,}
        Le~Liang,~\IEEEmembership{Member,~IEEE,}
        Jing~Zhang,~\IEEEmembership{Member,~IEEE,}
        Peiwen~Jiang,~\IEEEmembership{Graduate Student Member,~IEEE,}
        Yong~Li,~\IEEEmembership{Member,~IEEE,}
        and~Shi~Jin,~\IEEEmembership{Fellow,~IEEE}
\thanks{X.~Zhou, L. Liang, J.~Zhang, P.~Jiang, and S.~Jin are with the National Mobile
Communications Research Laboratory, Southeast University, Nanjing 210096, China
(e-mail: \protect \url{xy_zhou@seu.edu.cn}; lliang@seu.edu.cn; jingzhang@seu.edu.cn; peiwenjiang@seu.edu.cn; jinshi@seu.edu.cn). 
L. Liang is also with Purple Mountain Laboratories, Nanjing 211111, China.
S. Jin is also with the National Key Laboratory of Advanced Communication
Networks, Hebei Shijiazhuang 050081, China.}
\thanks{Y. Li is with the Academy of Network \& Communications of CETC, Hebei Shijiazhuang 050081, China, and also with the National Key Laboratory of Advanced Communication Networks, Hebei Shijiazhuang 050081, China
(e-mail:  \protect \url{young_li_54@126.com}).}
}

%
%

\maketitle

\begin{abstract}
Along with the prosperity of generative artificial intelligence (AI), its potential for solving conventional challenges in wireless communications has also surfaced. 
Inspired by this trend, we investigate the application of the advanced diffusion models (DMs), a representative class of generative AI models, to high dimensional wireless channel estimation. By capturing the structure of multiple-input multiple-output (MIMO) wireless channels via a deep generative prior encoded by DMs, we develop a novel posterior inference method for channel reconstruction. 
We further adapt the proposed method to recover channel information from low-resolution quantized measurements. 
Additionally, to enhance the over-the-air viability, we integrate the DM with the unsupervised Stein's unbiased risk estimator to enable learning from noisy observations and circumvent the requirements for ground truth channel data that is hardly available in practice.  
Results reveal that the proposed estimator achieves high-fidelity channel recovery while reducing estimation latency by a factor of 10 compared to state-of-the-art schemes, facilitating real-time implementation. 
Moreover, our method outperforms existing estimators while reducing the pilot overhead by half, showcasing its scalability to ultra-massive antenna arrays. 
\end{abstract}


%
\IEEEpeerreviewmaketitle

\section{Introduction}

\IEEEPARstart{T}o support the exploding demands of data transmission and diverse vertical application scenarios, next-generation wireless networks are poised to usher in an era characterized by tens of thousands of antennas and the use of terahertz frequency bands \cite{wang2024tutorial,sarieddeen2021overview}. In this context, channel estimation for the underlying high-dimensional multiple-input multiple-output (MIMO) systems is crucial for harnessing the potential gains of multiple antennas \cite{liang2014low}. For this purpose, the number of pilot symbols is conventionally assumed to exceed the number of transmit antennas, a requirement that becomes increasingly burdensome as antenna array sizes escalate.
Therefore, the challenge of acquiring accurate channel information with limited pilot overhead has garnered substantial research interest.

Traditional linear channel estimators, including the least squares (LS) and linear minimum mean-squared error (LMMSE) methods, are widely adopted but struggle to scale effectively to high-dimensional MIMO systems.
The LS estimator relies on sufficient pilot measurements no less than the transmit antenna counts, leading to significant training overhead.
For jointly Gaussian distributed channels and received signals, the LMMSE method gives the maximum \post estimate and generally achieves superior accuracy over LS \cite{kay1993fundamentals}. However, this estimator requires the statistics of channel correlation, typically demanding channel sample sizes proportional to the dimension of the covariance matrix.

A majority of existing works exploit the sparsity of high-dimensional wireless channels to simplify channel estimation using compressed sensing (CS)-based methods \cite{bajwa2010compressed}, assuming a sparse or low-rank channel representation in the angular domain or beamspace. Consequently, channel coefficients can be recovered with a reduced number of pilot measurements. 
Standard $l_1$-norm regularization in the angular domain was employed in \cite{schniterChannelEstimationPrecoder2014} to develop the LASSO method for millimeter wave (mmWave) channel estimation.
An alternative research direction investigated the application of approximate message passing (AMP) \cite{donohoMessagepassingAlgorithmsCompressed2009} for CS-based channel recovery.  
Specifically, AMP has been combined with the expectation maximization (EM) algorithm to achieve enhanced estimation performance over traditional CS methods \cite{vilaExpectationMaximizationGaussianMixtureApproximate2013,wen2014channel}. This strategy also found successful applications in systems with low-resolution analog-to-digital converters (ADCs) \cite{mo2017channel}.
Despite the wide variety of CS-based channel estimation methods being proposed, a significant roadblock before practical use is that the sparsity assumption would not exactly hold in realistic propagation scenarios. In other words, sparsity alone is insufficient to capture the structure of real-world channels, as this simple hand-crafted prior (or assumed channel model) fails to adequately represent the propagation environments across an entire urban cell \cite{fesl2024channel}.  

Following the trend of artificial intelligence (AI)-enhanced air interface, data-driven deep learning (DL) has been applied to address the dilemma of high-dimensional channel estimation \cite{heDeepLearningBasedChannel2018,chunDeepLearningBasedChannel2019,soltani2019deep}.
These studies generally involve neural networks that are trained in a supervised manner, where the network learns to map the received pilot signals to the estimated channels. 
In \cite{heDeepLearningBasedChannel2018}, a learned denoising-based AMP (LDAMP) algorithm was proposed, which incorporated a denoising convolutional neural network (CNN) into the iterative recovery structure of AMP, constructing a high-performance DL-based estimator.
Similarly, a two-stage DL-based estimation scheme was developed in \cite{chunDeepLearningBasedChannel2019}, integrating pilot design and data-aided strategies for enhancement.
However, these supervised DL-based methods are measurement-specific and do not generalize effectively when the measurement setups, such as pilot and antenna counts, are altered.

As a representative technology in the new era of AI, generative AI has been revolutionizing the forefronts of both academia and industry.  
Deep generative modeling 
has shown potent abilities in capturing highly complex relationships, facilitating high-dimensional data synthesis and efficient data transmission, and circumventing the curse of dimensionality \cite{ye2020deep,jiang2023semantic}. 
Particularly, generative diffusion models (DMs) \cite{sohl2015deep,ho2020denoising,song2020score} have significantly advanced generative modeling due to their ability to produce diverse high-quality samples through simple training implementations, laying the groundwork for numerous popular generative foundation models \cite{rombach2022high}.

Leveraging deep generative modeling to address conventional challenges in wireless transmission has also emerged as a prominent topic.
For channel estimation, generative AI models such as generative adversarial networks (GANs) \cite{baleviHighDimensionalChannel2021,balevi2021wideband}, variational autoencoders \cite{baurLeveragingVariationalAutoencoders2023}, and Gaussian mixture models \cite{koller2022asymptotically}, have been utilized for performance enhancement. These deep generative networks capture prior knowledge and offer a model-agnostic approach to characterize inherently structured or sparse channels, circumventing channel model assumptions that may not be applicable.

As the most popular generative model, DMs' proficiency in modeling intricate distributions and furnishing robust prior information underscores its promise in tackling inverse problems such as channel estimation.
A MIMO channel estimator utilizing score-based generative models (SGMs), closely associated with DMs, was proposed in \cite{arvinteMIMOChannelEstimation2022}, achieving state-of-the-art estimation accuracy and robust out-of-distribution performance. 
However, the practical value of this scheme is hindered by the huge number of inference steps and network parameters involved, hardly satisfying the latency requirement of channel estimation while leading to a substantial storage burden.  
To address the challenge of real-time implementation, the authors of \cite{fesl2024diffusion} develop a low-complexity DM-based channel estimator using a lightweight CNN, which learns the channel prior within the angular domain. However, the proposed method necessitates orthogonal pilot sequences and full pilot measurements to conduct an LS estimate as initialization, leading to substantial overhead in high-dimensional MIMO systems. 
Moreover, applying these DM-based estimators \cite{arvinteMIMOChannelEstimation2022,fesl2024diffusion} to quantized MIMO receivers with low-resolution ADCs remains an open issue. 
Additionally, existing works assume the availability of a large dataset containing clean channel samples for training DMs. However, collecting such a dataset can be difficult or even impossible for realistic air interfaces.

In this paper, we propose a DM-based posterior inference method for high-dimensional MIMO channel estimation. Distinct from existing schemes, we design a conditional posterior sampling process as the per-step update rule to recover the original channel from noise. The proposed method achieves precise estimation in the low-pilot regime with significantly lower latency than state-of-the-art schemes, 
while maintaining favorable scalability to various system setups. Moreover, we apply our approach to channel estimation with low-resolution ADCs and scenarios wherein only noisy training data is available.
The contributions of this paper are summarized as follows.

\begin{itemize}
  
  \item \textbf{DM-Based Channel Estimation:} We utilize the powerful DM as a data-driven generative prior to characterize high-dimensional MIMO channels and formulate a posterior channel estimator given pilot observations. By combining prior information from the pre-trained DM with a closed-form approximation of the likelihood term, our method iteratively conducts conditional posterior estimation at each step. This posterior inference method, combined with a lightweight network architecture for the DM, enables high-fidelity estimation with reduced latency.  

  \item \textbf{Applying to Quantized Channel Estimation:} We adapt the proposed method to channel estimation with low-resolution ADCs by modifying the likelihood information corresponding to the quantized measurements. To the best of our knowledge, this is the first work to investigate the applications of DMs in  
  few-bit quantized receivers. The developed scheme achieves notable performance enhancements over state-of-the-art quantized channel estimators while maintaining low latency. 

  \item \textbf{Learning from Noisy Channel Realizations:} We integrate Stein's unbiased risk estimator (SURE) denoising \cite{stein1981estimation} into the training of DMs to enable learning with noisy channel data. 
  Results show that the model learned via the proposed scheme provides robust prior knowledge for channel estimation and enables competitive recovery accuracy. 
  This strategy alleviates the requirements for substantial clean channel datasets and facilitates over-the-air (OTA) implementation.
\end{itemize}

\textit{Notations:} 
For any matrix $\mathbf{A}$, $\mathbf{A}^T$, $\mathbf{A}^{\rm H}$, and $\mathbf{A}^{-1}$ represent the transpose, conjugate transpose, and inverse of $\mathbf{A}$, respectively. 
Also, 
$\mathbf{0}$ is a zero vector, 
$\mathbf{I}$ is an identity matrix, 
$\|\cdot\|_2$ denotes the $l_2$-norm, 
$\vector(\cdot)$ denotes the vectorization operation,  
$\otimes$ denotes the Kronecker product, 
$\mathbb{E}[\cdot]$ denotes the expectation operator, and 
$\mathbb{I}(\cdot)$ represents an indicator function that returns value one only when the event as the argument is true and zero otherwise. Moreover,
$\re(\cdot)$ and $\im(\cdot)$ extract the real and imaginary parts of a complex value, respectively. 
$\mathcal{N}(z;\mu,\sigma^2)$ represents a Gaussian random variable $z$ with mean $\mu$ and variance $\sigma^2$. 
Finally, 
$\mathbb{C}$ and $\mathbb{R}$ denote the sets of complex and real numbers, respectively,  
$[N]$ is the set of nonnegative integers up to $N$, and
$\mathcal{U}([N])$ denotes a discrete uniform distribution within $[N]$.

\section{Problem Formulation and Preliminaries}

This section commences with the formulation of the MIMO channel estimation problem. 
Subsequently, a brief overview of DMs is provided as the preliminaries. 

\subsection{MIMO Channel Estimation} 
Consider a point-to-point, narrowband MIMO system with $\nt$ transmit and $\nr$ receive antennas. A total of $\np$ symbols are allocated for pilot transmissions and channel estimation, where the $k$-th pilot symbol is denoted as $\mathbf{p}_k\in \mathbb{C}^{\nt \times 1}, k\in[\np]$. The received signal at this $k$-th slot is given by
\CheckRmv{
  \begin{equation}
    \mathbf{y}_k = \mathbf{Hp}_k+\mathbf{n}_k,
  \end{equation}
}
where $\mathbf{H}\in \mathbb{C}^{\nr \times \nt}$ represents the MIMO channel matrix to be estimated, and $\mathbf{n}_k\sim \mathcal{CN}(\mathbf{0}, 2\sigma_n^2\mathbf{I})$ denotes the additive white Gaussian noise (AWGN) vector with variance $\sigma_n^2$ per real/imaginary component.
The MIMO channels are assumed to be quasi-static, remaining constant within the $\np$ pilot transmissions. Therefore, the received signal during this training period can be written in the matrix form $\mathbf{Y}\in\mathbb{C}^{\nr\times\np}$ as\footnote{{Although we assume a fully digital architecture in the system model, the proposed method can be readily extended to massive MIMO systems with hybrid analog-digital architectures.}}
\CheckRmv{
  \begin{equation}
    \mathbf{Y} = \mathbf{HP} + \mathbf{N}, 
    \label{eq:model_matrix}
  \end{equation}
}
where $\mathbf{P}=[\mathbf{p}_1,\ldots,\mathbf{p}_{\np}]$ and $\mathbf{N}=[\mathbf{n}_1,\ldots,\mathbf{n}_{\np}]$ contain the stacked pilot symbols and noise vectors, respectively. 
For simplicity, the pilot symbols are supposed to be equiprobably drawn from the quadrature phase shift keying (QPSK) lattice, i.e., $\{\pm 1/\sqrt{2} \pm {\rm j}/\sqrt{2}\}$, fulfilling the unit power constraint. 

Using vectorization, the signal model \eqref{eq:model_matrix} can be reformulated as
\CheckRmv{
  \begin{equation}
    \vec{\mathbf{y}} = \bar{\mathbf{A}}\vec{\mathbf{h}} + \vec{\mathbf{n}}, 
    \label{eq:model_vec_cplx}
  \end{equation}
}
where $\vec{\mathbf{y}}=\vector(\mathbf{Y})$, $\vec{\mathbf{h}}=\vector(\mathbf{H})$, $\vec{\mathbf{n}}=\vector(\mathbf{N})$, and $\bar{\mathbf{A}} = \mathbf{P}^T \otimes \mathbf{I}_{\nr}$.
The MIMO channels are normalized to ensure that $\mathbb{E}[|\vec{h}_{ij}|^2] = 1$. 
The signal-to-noise power ratio (SNR) is defined as $\nt / (2\sigma_n^2)$.

Consider the angular domain representation of MIMO channels, which is a natural choice for channels at high-frequency bands characterized by high dimensionality and pronounced propagation directionality \cite{bajwa2010compressed}.
Assuming the use of uniform linear arrays (ULAs) with half-wavelength spacing at both the transmitter and receiver, the so-called virtual channel model \cite{sayeed2002deconstructing} allows us to represent $\vec{\mathbf{h}}$ with respect to the angular domain channel $\vec{\mathbf{h}}_{\rm ad}$ as 
\CheckRmv{
  \begin{equation}
     \vec{\mathbf{h}} = \big((\mathbf{A}_{\rm T}^{T})^{\rm H} \otimes \mathbf{A}_{\rm R}\big) \vec{\mathbf{h}}_{\rm ad},
     \label{eq:angle}
  \end{equation}
}
where $\mathbf{A}_{\rm T}\in \mathbb{C}^{\nt \times \nt}$ and $\mathbf{A}_{\rm R} \in \mathbb{C}^{\nr \times \nr}$ denote the discrete Fourier transform matrices corresponding to the array response matrices at the transmitter and receiver, respectively. 
Substituting \eqref{eq:angle} into \eqref{eq:model_vec_cplx}, we derive 
\CheckRmv{
  \begin{equation}
    \vec{\mathbf{y}} = {\mathbf{A}}_{\rm ad}\vec{\mathbf{h}}_{\rm ad} + \vec{\mathbf{n}}, 
    \label{eq:model_vec_angle}
  \end{equation}
}
where ${\mathbf{A}}_{\rm ad} = \bar{\mathbf{A}}\big((\mathbf{A}_{\rm T}^{T})^{\rm H} \otimes \mathbf{A}_{\rm R}\big) \in \mathbb{C}^{\nr\np \times \nr\nt}$. 
For notational convenience, we consider the equivalent real-valued form of \eqref{eq:model_vec_angle} throughout this paper, given as
\CheckRmv{
  \begin{equation}
    \mathbf{y} = \mathbf{Ah} + \mathbf{n}, 
    \label{eq:model_vec}
  \end{equation}
}
where $\mathbf{y}\in \mathbb{R}^{M\times 1}$, $\mathbf{A}\in \mathbb{R}^{M\times N}$, $\mathbf{h}\in \mathbb{R}^{N\times 1}$, and $\mathbf{n}\in \mathbb{R}^{M\times 1}$, 
with $M=2\nr\np$ and $N=2\nr\nt$.

The objective of channel estimation is to estimate $\mathbf{h}$ 
given the received pilot measurements ${\mathbf{y}}$ and the knowledge of the pilot $\mathbf{P}$. 
Let $\alpha=\np/\nt$ denote the pilot density. When $\alpha<1$, channel estimation becomes an under-determined linear inverse problem, requiring strong priors for regularization.\footnote{{In addition to leveraging strong prior information, such as deep generative priors, to reduce pilot overhead, deep learning-based channel extrapolation has also emerged as an effective alternative strategy \cite{zhang2021deep}. However, as this paper focuses on a fundamentally different approach, the detailed mechanisms of channel extrapolation and its comparison with the proposed method are beyond the scope of this work.}}

\subsection{Denoising Diffusion Generative Models} \label{sec:dm} 
DMs are among the most advanced generative learning models. This class of models establishes a Markov chain of diffusion steps to incrementally introduce random noise to data and subsequently learn to invert the forward process to enable high-quality data sample generation from noise \cite{sohl2015deep,ho2020denoising}.
Next, we introduce DMs with vectorized notations for convenience.  
Specifically, given a data prior distribution $\mathbf{h}_0\sim p_0(\cdot)$, DMs' forward process adds noise using a Gaussian transition kernel for $T$ time steps,
\CheckRmv{
  \begin{equation}
    p(\mathbf{h}_t |\mathbf{h}_{t-1}) = \mathcal{N}(\mathbf{h}_t;\sqrt{1-\beta_t}\mathbf{h}_{t-1},\beta_t\mathbf{I}),\;t\in [T],
    \label{eq:ht-1_to_ht}
  \end{equation}
}
where $\mathbf{h}_t$ is the latent variable at time step $t$, and $0<\beta_1<\beta_2<\ldots<\beta_T<1$ denote the prescribed noise schedule. 
Defining $\alpha_t\triangleq 1-\beta_t$ and $\bar{\alpha}_t\triangleq\prod_{i=1}^t \alpha_i$ and using the reparameterization trick, 
it is easy to derive that
\CheckRmv{
  \begin{equation}
    \mathbf{h}_t = \sqrt{\bar{\alpha}_t}\mathbf{h}_0 + \sqrt{1-\bar{\alpha}_t}\boldsymbol{\epsilon}_t,\; \boldsymbol{\epsilon}_t\sim\mathcal{N}(\mathbf{0},\mathbf{I}).  \label{eq:h0_to_ht}
  \end{equation}
}
The noise schedule is crafted such that $\bar{\alpha}_T\approx 0$, and $\mathbf{h}_T$ approximately follows an isotropic Gaussian distribution $\mathcal{N}(\mathbf{0},\mathbf{I})$.

Ideally, the reverse process recreates data samples from pure Gaussian noise by sampling from the reverse conditional distribution $p(\mathbf{h}_{t-1} | \mathbf{h}_{t})$. However, $p(\mathbf{h}_{t-1} | \mathbf{h}_{t})$ is generally intractable to calculate due to the unknown prior $p_0(\cdot)$. 
Hence, DMs learn a variational Markov chain $p_{\boldsymbol{\theta}}(\mathbf{h}_{t-1} | \mathbf{h}_{t})$ parameterized by $\boldsymbol{\theta}$ to conduct the reverse process $\mathbf{h}_T\to\mathbf{h}_{T-1}\to\cdots\to \mathbf{h}_0$.
Note that the reverse conditional distribution is tractable if further conditioned on $\mathbf{h}_0$ \cite{ho2020denoising}, which also becomes Gaussian,  
\CheckRmv{
  \begin{align}
    p(\mathbf{h}_{t-1} | \mathbf{h}_{t}, \mathbf{h}_0) &= \mathcal{N}(\mathbf{h}_{t-1} ; \tilde{\boldsymbol{\mu}}_t, \tilde{\beta}_{t} \mathbf{I}), \\
    \tilde{\boldsymbol{\mu}}_t&=\frac{1}{\sqrt{\alpha_t}}\left(\mathbf{h}_t-\frac{1-\alpha_t}{\sqrt{1-\bar{\alpha}_t}}\boldsymbol{\epsilon}_t\right),\label{eq:tilde_mu}\\
    \tilde{\beta}_{t} &= \frac{1-\bar{\alpha}_{t-1}}{1-\bar{\alpha}_{t}}\beta_t.
  \end{align}
}
Therefore, it is reasonable to parameterize $p_{\boldsymbol{\theta}}$ as Gaussian,
given by $\mathcal{N}(\mathbf{h}_{t-1} ; \boldsymbol{\mu}_{\boldsymbol{\theta}}(\mathbf{h}_t,t), \tilde{\beta}_{t} \mathbf{I})$, where $\boldsymbol{\mu}_{\boldsymbol{\theta}}$ is used to approximate $\tilde{\boldsymbol{\mu}}_t$.
The objective is to learn an approximation of $p(\mathbf{h}_{t-1} | \mathbf{h}_{t}, \mathbf{h}_0)$ averaged across $\mathbf{h}_0$, which amounts to minimizing a series of Kullback-Leibler divergences between Gaussians \cite{ho2020denoising}. 

The authors of \cite{ho2020denoising} further parameterize $\boldsymbol{\mu}_{\boldsymbol{\theta}}$ using the same functional form as in \eqref{eq:tilde_mu},  replacing $\boldsymbol{\epsilon}_t$ with a neural network $\boldsymbol{\epsilon}_{\boldsymbol{\theta}}(\mathbf{h}_t,t)$. 
This network, referred to as the denoising network throughout this paper, takes the noisy $\mathbf{h}_t$ in \eqref{eq:h0_to_ht} and the time step $t$ as inputs and predicts the additive noise $\boldsymbol{\epsilon}_t$ for each time step.
Consequently, a practically effective loss function is identified as follows
\CheckRmv{
  \begin{equation}
    \mathcal{L}_{\rm DM}(\boldsymbol{\theta}) = \mathbb{E}_{\mathbf{h}_0,\boldsymbol{\epsilon}_t,t}
    \left[{{\left\| {\boldsymbol{\epsilon}_t - {\boldsymbol{\boldsymbol{\epsilon}}_{\boldsymbol{\theta}}}(\sqrt{\bar{\alpha}_t}\mathbf{h}_0 + \sqrt{1-\bar{\alpha}_t}\boldsymbol{\epsilon}_t,t)} \right\|_2^2}} \right],
    \label{eq:dm_loss}
  \end{equation}
}
where $t \sim {\mathcal{U}}([T])$.
After training, the parameters of $\boldsymbol{\epsilon}_{\boldsymbol{\theta}}$ are fixed, and data samples are generated by the following iterative step for $t=T$ to 1, 
\CheckRmv{
  \begin{equation}
    \mathbf{h}_{t-1} = \frac{1}{\sqrt{\alpha_t}}\left(\mathbf{h}_t-\frac{1-\alpha_t}{\sqrt{1-\bar{\alpha}_t}}\boldsymbol{\epsilon}_{\boldsymbol{\theta}}(\mathbf{h}_t,t)\right) + \tilde{\beta}_{t}\mathbf{z}_t, 
    \label{eq:reverse_dm}
  \end{equation}
}
where $\mathbf{z}_t\sim\mathcal{N}(\mathbf{0},\mathbf{I})$.

\section{Proposed Methods} 
First, we establish the DM-based channel estimation scheme targeting the full-resolution system \eqref{eq:model_vec}.  
Subsequently, we investigate the application to scenarios with low-resolution ADCs.
Finally, we consider training adaptations that enable learning the DM without ground truth channel data to enhance the practical viability of the proposed method. 

\subsection{Diffusion Model-Based Channel Estimation}

\CheckRmv{
  \begin{figure}[t]
    \centering
    \includegraphics[width=3.45in]{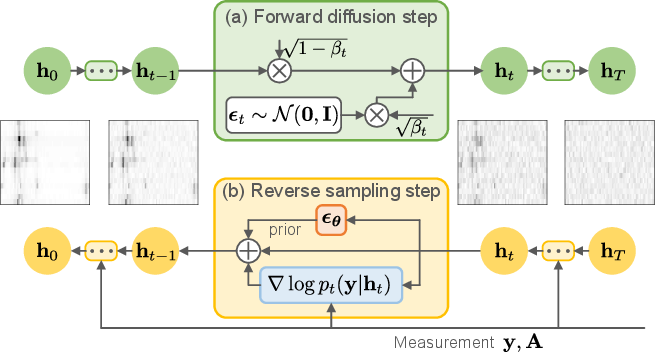}
    \caption{Block diagram of the proposed method, where (a) illustrates the forward diffusion involved in the training phase, and (b) depicts the reverse sampling step in the inference phase.}
    \label{fig:diagram}
  \end{figure}
}

We develop a posterior inference method for channel estimation, leveraging the DM as a deep generative prior. According to the Bayesian philosophy, the channel estimation problem of recovering $\mathbf{h}$ from $\mathbf{y}$ in \eqref{eq:model_vec} can be framed as conducting posterior inference on $\mathbf{h}$, targeting the posterior distribution $p(\mathbf{h}|\mathbf{y}) \propto p_0(\mathbf{h})p(\mathbf{y}|\mathbf{h})$. 
We decouple this task into learning the unknown channel prior $p_0(\cdot)$ in the offline training phase 
and recovering channels using the learned prior and pilot measurements in the online inference phase.
The block diagram of the proposed method is illustrated in \figref{fig:diagram}.

\subsubsection{Learning a DM for Wireless Channels}
The prior channel distribution $p_0(\cdot)$ is generally hand-crafted in CS-based methods, such as the $l_0$ sparsity used in orthogonal matching pursuit \cite{alkhateeb2014channel} and the $l_1$ sparsity used in LASSO \cite{schniterChannelEstimationPrecoder2014}. 
However, these hand-crafted priors may not fully represent the intricate structure of realistic wireless channels.  
To address this challenge, a pre-trained DM can be utilized as a data-driven prior by providing the score function\footnote{The score function of a distribution $p(\mathbf{h})$ is given by the gradient of the log-probability, i.e., $\nabla_{\mathbf{h}} \log p(\mathbf{h})$, which enables sampling from $p(\mathbf{h})$ using Langevin dynamics \cite{song2020score,vincent2011connection,luo2022understanding}.} of the underlying channel distribution without assuming specific structures of the distribution. 
We follow \cite{fesl2024diffusion} to adopt a lightweight CNN architecture (\figref{fig:network} in Appendix~\ref{appendix:net}) for the DM's denoising network $\boldsymbol{\epsilon}_{\boldsymbol{\theta}}$ and learn the channel distribution in the angular domain. This virtual channel representation is known to be compressible \cite{rappaport2019wireless}, especially for high-dimensional channels, such that the network parameters can be significantly diminished.  
Note that this strategy does not impose any assumptions on the structure or sparsity of the channel distribution.  

The denoising network $\boldsymbol{\epsilon}_{\boldsymbol{\theta}}$ takes $\mathbf{h}_t$ as the input, with the Transformer sinusoidal positional embedding \cite{vaswani2017attention} specifying the time step index $t$, utilizes several convolutional layers for feature extraction,   and finally outputs the predicted noise. Details of the network structure can be found in Appendix~\ref{appendix:net}.  
The network is trained by minimizing {the mini-batch version} of the loss function 
in \eqref{eq:dm_loss}. 
At each training step, a random data sample $\mathbf{h}_0$ is drawn from the channel dataset. Concurrently, random noise $\boldsymbol{\epsilon}\sim \mathcal{N}(\mathbf{0}, \mathbf{I})$ is generated and added to the selected sample $\mathbf{h}_0$ according to \eqref{eq:h0_to_ht}, where the ratio $\bar{\alpha}_t$ is determined by a time step index $t$ uniformly chosen from $[T]$. 
Then, the network parameters $\boldsymbol{\theta}$ can be updated by taking the gradient descent step on
\CheckRmv{
  \begin{equation}
    \nabla_{\boldsymbol{\theta}}\left\| {\boldsymbol{\epsilon} - {\boldsymbol{\boldsymbol{\epsilon}}_{\boldsymbol{\theta}}}(\sqrt{\bar{\alpha}_t}\mathbf{h}_0 + \sqrt{1-\bar{\alpha}_t}\boldsymbol{\epsilon},t)} \right\|_2^2.
  \end{equation}
}
It is noteworthy that no information on the measurement model, e.g., the measurement matrix $\mathbf{A}$ and noise statistics, is involved in the training process, differentiating our approach from supervised learning-based channel estimators. This measurement model-agnostic DM-based generative approach 
enhances robustness against variations in pilot counts, noise distributions, and SNR values.

\subsubsection{Derivation of the Posterior Inference Algorithm}
Our approach to solving the inverse problem in \eqref{eq:model_vec} originates from using the score of the posterior distribution for diffusion posterior sampling \cite{song2020score}.
Since DM's forward process gradually perturbs the data with noise, what is required 
is the \textit{noise-perturbed posteriors score} $\nabla_{\mathbf{h}_t} \log p_t(\mathbf{h}_t|\mathbf{y})$ for $\mathbf{h}_t$ with varying noise levels \cite{meng2022diffusion}.\footnote{Note that we use the index $t$ to distinguish the noise-perturbed distribution at each time step from the distribution of the original clean data.}
Aligning with the reverse process of the DM, we use an iterative algorithm for channel estimation, where each step $t$ (from $t=T$ to 1) applies the update rule as follows
\CheckRmv{ 
  \begin{equation}
   {\mathbf{h}}_{t-1}=\frac{1}{\sqrt{\alpha_t}}\big(\mathbf{h}_t + (1-{\alpha}_t)\nabla_{\mathbf{h}_t} \log p_t(\mathbf{h}_t|\mathbf{y})\big). 
    \label{eq:condition_post}
  \end{equation}
}
The intuition to employ this update rule is supported by the following proposition, proved in Appendix~\ref{appendix:theorem}.

\begin{proposition}
  The update rule in \eqref{eq:condition_post} gives a posterior mean estimate of the latent variable   
  $\mathbf{h}_{t-1}$ conditioned on $\mathbf{h}_t$ and $\mathbf{y}$, i.e., $\mathbb{E}[\mathbf{h}_{t-1}|\mathbf{h}_t,\mathbf{y}]=\int \mathbf{h}_{t-1} p_{t-1}(\mathbf{h}_{t-1}|\mathbf{h}_{t},\mathbf{y}){\rm d}\mathbf{h}_{t-1}$. 
  \label{th:posterior_mean}
\end{proposition}  

{\begin{remark}
  As indicated by Proposition~\ref{th:posterior_mean}, the conditional posterior mean update outlined in \eqref{eq:condition_post} utilizes the posterior information, which is a combination of the prior and likelihood, to satisfy both prior and measurement consistency of the estimate. This leads to samples in high probability regions of the channel posterior distribution, particularly as $t \to 0$, boosting estimation performance. 
  Moreover, the update rule follows the deterministic sampling paradigm \cite{song2020denoising}, eliminating the extra perturbation added 
  in the reverse process of DM ($\mathbf{z}_t$ in \eqref{eq:reverse_dm}). This strategy is known to enhance sample quality when fewer time steps are utilized.
\end{remark}}

To derive the posterior score, the crucial observation is that, by applying Bayes' rule, it can be decomposed as \cite{jalal2021robust,meng2022diffusion} 
\CheckRmv{
  \begin{equation}
    \nabla_{\mathbf{h}_t} \log p_t(\mathbf{h}_t|\mathbf{y}) = \nabla_{\mathbf{h}_t} \log p_t(\mathbf{h}_t) + \nabla_{\mathbf{h}_t} \log p_t(\mathbf{y}|\mathbf{h}_t),
  \end{equation}
}
where the corresponding {prior score} can be approximated by the pre-trained DM's denoising network as follows \cite{song2020score,luo2022understanding}
\CheckRmv{
  \begin{equation}
    \nabla_{\mathbf{h}_t} \log p_t(\mathbf{h}_t) \approx - \frac{1}{\sqrt{1-\bar{\alpha}_t}}\boldsymbol{\epsilon}_{\boldsymbol{\theta}}(\mathbf{h}_t,t),
    \label{eq:score_dm}
  \end{equation}
} 
where $p_t(\mathbf{h}_t) = \int {p_0(\mathbf{h}_0)p(\mathbf{h}_t|\mathbf{h}_0){\rm d}\mathbf{h}_0}$ represents the noise-perturbed data distribution at time step $t$. 
However, the likelihood score $\nabla_{\mathbf{h}_t} \log p_t(\mathbf{y}|\mathbf{h}_t)$ becomes intractable for $t>0$ \cite{meng2022diffusion,song2022pseudoinverse}. 
To illustrate this assertion, let us revisit the definition of DM, which allows for representing $p_t(\mathbf{y}|\mathbf{h}_t)$ as $p_t(\mathbf{y}|\mathbf{h}_t) = \int {p(\mathbf{y}|\mathbf{h}_0)p(\mathbf{h}_0|\mathbf{h}_t){\rm d}\mathbf{h}_0}$. Herein, we have used the fact that $\mathbf{h}_t$ and $\mathbf{y}$ are independent conditioned on $\mathbf{h}_0$, verified by the graphical model $\mathbf{y}\gets \mathbf{h}_0\to \mathbf{h}_t$ of the DM \cite{song2022pseudoinverse}. In the integral, 
the reverse probability $p(\mathbf{h}_0|\mathbf{h}_t)$ can only be approximated by sampling from the entire DM, as discussed in \secref{sec:dm}, making it challenging to compute.

To resolve the challenge in handling $\nabla_{\mathbf{h}_t} \log p_t(\mathbf{y}|\mathbf{h}_t)$, we resort to approximations using an uninformative prior assumption inspired by \cite{meng2022diffusion}. Specifically, by assuming $p_0(\mathbf{h}_0)$ is uninformative with respect to $p(\mathbf{h}_t|\mathbf{h}_0)$, we have $p(\mathbf{h}_0|\mathbf{h}_t) \propto p(\mathbf{h}_t|\mathbf{h}_0)p_0(\mathbf{h}_0)\approx p(\mathbf{h}_t|\mathbf{h}_0)$. This assumption is asymptotically precise when $t\to 0$, as shown in the exemplified verification in \cite[Appendix A]{meng2022diffusion}. To see why it works, recall that from \eqref{eq:h0_to_ht} we have
\CheckRmv{
  \begin{equation}
    p(\mathbf{h}_t|\mathbf{h}_0) \propto \exp \left(-\frac{\|\mathbf{h}_t - \sqrt{\bar{\alpha}_t}\mathbf{h}_0\|_2^2}{2(1-\bar{\alpha}_t)}\right),
  \end{equation}
} 
which notably increases as $t\to 0$ due to the small amount of added noise ($\bar{\alpha}_t\to 1$) at the initial stage of the forward process. Hence, the conditional probability $p(\mathbf{h}_t|\mathbf{h}_0)$ dominates, and ignoring the prior would not result in significant deviations. 

Based on the uninformative prior assumption, the noise-perturbed likelihood score can be derived in closed form as follows. Owing to the approximation $p(\mathbf{h}_0|\mathbf{h}_t) \propto p(\mathbf{h}_t|\mathbf{h}_0) \approx \mathcal{N}(\mathbf{h}_0;\frac{1}{\sqrt{\bar{\alpha}_t}}\mathbf{h}_t, \frac{1-\bar{\alpha}_t}{\bar{\alpha}_t}\mathbf{I})$, $\mathbf{h}_0$ can be equivalently expressed as 
\CheckRmv{
  \begin{equation}
    \mathbf{h}_0 = \frac{1}{\sqrt{\bar{\alpha}_t}}\left(\mathbf{h}_t + \sqrt{1-\bar{\alpha}_t}\mathbf{u}\right),\;\mathbf{u}\sim\mathcal{N}(\mathbf{0},\mathbf{I}).
    \label{eq:pseudo_likelihood_h}
  \end{equation}
}
Therefore, the received signal $\mathbf{y}$ in \eqref{eq:model_vec} can be alternatively represented as
\CheckRmv{
  \begin{equation}
    {\mathbf{y}}=\frac{1}{\sqrt{\bar{\alpha}_t}}\mathbf{A}\mathbf{h}_t + \frac{\sqrt{1-\bar{\alpha}_t}}{\sqrt{\bar{\alpha}_t}}\mathbf{A}\mathbf{u} + \mathbf{n}.
    \label{eq:ht_to_y}
  \end{equation}
}
Therefore, the noise-perturbed likelihood is in the form of a Gaussian distribution given by
\CheckRmv{
  \begin{equation}
    p_{t}(\mathbf{y}|\mathbf{h}_{t})=\mathcal{N}\left(\mathbf{y};\frac{1}{\sqrt{\bar{\alpha}_{t}}}\mathbf{A}\mathbf{h}_{t},\frac{1-\bar{\alpha}_{t}}{\bar{\alpha}_{t}}\mathbf{A}\mathbf{A}^T+\sigma_{n}^{2}\mathbf{I}\right),
  \end{equation}
}
and its score can be written as
\CheckRmv{
  \begin{align}
    &\nabla_{\mathbf{h}_t}\log p_t(\mathbf{y}|\mathbf{h}_t)\nonumber\\
    &=\frac{1}{\sqrt{\bar{\alpha}_t}}\mathbf{A}^T\left(\frac{1-\bar{\alpha}_t}{\bar{\alpha}_t}\mathbf{A}\mathbf{A}^T+\sigma_n^2\mathbf{I}\right)^{-1}\Big(\mathbf{y}-\frac{1}{\sqrt{\bar{\alpha}_t}}\mathbf{A}\mathbf{h}_t\Big).
    \label{eq:pseudo_likelihood_inv}
  \end{align}
}
\begin{remark}
  To mitigate the high-complexity matrix inversion in \eqref{eq:pseudo_likelihood_inv}, based on the singular value decomposition (SVD) $\mathbf{A}=\mathbf{U}\boldsymbol{\Sigma}\mathbf{V}^T$, we can reformulate the computation as follows
  \CheckRmv{
    \begin{align}
      &\nabla_{\mathbf{h}_t}\log p_t(\mathbf{y}|\mathbf{h}_t)\nonumber\\
      &=\frac{1}{\sqrt{\bar{\alpha}_{t}}}\mathbf{V}\mathbf{\Sigma}\left(\frac{1-\bar{\alpha}_{t}}{\bar{\alpha}_{t}}\mathbf{\Sigma}^{2}+\sigma_n^{2}\mathbf{I}\right)^{-1}\Big(\mathbf{U}^T\mathbf{y}-\frac{1}{\sqrt{\bar{\alpha}_{t}}}\mathbf{\Sigma}\mathbf{V}^T\mathbf{h}_{t}\Big).
      \label{eq:pseudo_likelihood_svd}
    \end{align} 
  }
  The SVD can be computed in advance and reused as long as the measurement matrix $\mathbf{A}$ remains fixed. Hence, significant computational costs are saved as compared to the per-step matrix inversion in \eqref{eq:pseudo_likelihood_inv}.    
\end{remark}

\renewcommand{\algorithmicrequire}{\textbf{Input:}}
\renewcommand{\algorithmicensure}{\textbf{Output:}}
\renewcommand{\algorithmiccomment}[1]{/* #1 */}
\newcommand{\IfThen}[2]{
  \STATE \algorithmicif\ #1\ \algorithmicthen\ #2}
\newcommand{\parfor}[1]{\STATE \algorithmicfor\ #1  \textbf{do in parallel}}
\CheckRmv{
  \begin{algorithm}[t]
    \caption{DM-Based Channel Estimation}
    \label{alg:dmce}
    \begin{algorithmic}[1] 
      \REQUIRE ${\bf{A}}$, ${\bf{y}}$, $\sigma_n^2$, pre-trained denoising network $\boldsymbol{\epsilon}_{\boldsymbol{\theta}}$, noise schedule $\{\beta_t\}_{t=1}^T$, gradient scale $s$. 
      \STATE \textbf{Initialize:} $\mathbf{h}_{T}\sim \mathcal{N}(\mathbf{0}, \mathbf{I})$.
      \STATE \textbf{Compute:} $\alpha_t=1-\beta_t, \; \bar{\alpha}_t = \prod_{i=1}^{t}\alpha_t,\; t\in[T]$.	
      \FOR{$t=T$ to $1$}
        \STATE $\mathbf{h}_{t-1}=\frac{1}{\sqrt{\alpha_t}}\left(\mathbf{h}_t-\frac{1-\alpha_t}{\sqrt{1-\bar{\alpha}_t}}\boldsymbol{\epsilon}_{\boldsymbol{\theta}}(\mathbf{h}_t,t)\right)$.
        \STATE Compute $\boldsymbol{l} = \nabla_{\mathbf{h}_t}\log p_t(\mathbf{y}|\mathbf{h}_t)$ using \eqref{eq:pseudo_likelihood_svd}.
        \STATE $\mathbf{h}_{t-1}\gets \mathbf{h}_{t-1} + s \frac{1-\alpha_t}{\sqrt{\alpha_t}} \boldsymbol{l}$.
      \ENDFOR
      \ENSURE Estimated channel $\hat{\mathbf{h}} = \mathbf{h}_{0}$. 
    \end{algorithmic} 
  \end{algorithm}
}

The proposed DM-based channel estimation scheme is outlined in Algorithm \ref{alg:dmce} and illustrated in the lower half of \figref{fig:diagram}. Line 4 of the algorithm encodes the deep generative prior into the update, and Line 6 further incorporates the likelihood score. According to  \cite{meng2022diffusion} and \cite{dhariwal2021diffusion}, we adopt a gradient scale parameter $s$ in the update rule to effectively weigh between the prior and likelihood. In literature, it is observed that $s>1$ yields empirically enhanced performance. This can be attributed to the fact that the scaled score $s\cdot \nabla_{\mathbf{h}_t}\log p_t(\mathbf{y}|\mathbf{h}_t)$ leads to a sharper likelihood $p_t(\mathbf{y}|\mathbf{h}_t)^s$ that emphasizes the mode of the posterior distribution, aiding in the generation of high-quality estimates \cite{dhariwal2021diffusion}. Nevertheless, our algorithm is robust to the selection of $s$, as shown in experiments.

\subsection{DM-Based Quantized Channel Estimation}
To reduce power consumption, low-resolution ADCs can be deployed at the receiver end. The quantized received signal is given by
\CheckRmv{
  \begin{equation}
    \bar{\mathbf{y}}=Q(\mathbf{y})=Q(\mathbf{Ah}+\mathbf{n}),
    \label{eq:model_quan}
  \end{equation}
}
where $Q(\cdot)$ denotes the quantization function of the ADCs. 
A scalar quantizer that performs element-wise on the input is assumed due to its simplicity and widespread usage. 

For ADCs with $b$ quantization bits, the quantizer maps the input into a countable set of codewords, $\mathcal{R} = \{r_1, r_2, \ldots, r_{2^b - 1}\}$. The codewords correspond to intervals split by the quantization thresholds $\mathcal{T}=\{(r_k^{\rm low}, r_k^{\rm up}) \mid k\in[2^b]\}$. For example, if $Q(z)=\bar{z} \in \mathcal{R}$, the input $z$ would satisfy $\bar{z}^{\rm low}\leq z < \bar{z}^{\rm up}$, where $\bar{z}^{\rm low}$ and $\bar{z}^{\rm up}$ represent the lower and upper thresholds with respect to the codeword $\bar{z}$, respectively.
Considering a uniform mid-rise quantizer\footnote{The proposed channel estimation method is not confined to this choice and can be readily extended to more complex non-uniform quantizers.}, we have $r_k = (2k - 2^b -1)\frac{\Delta}{2},\;k\in [2^b] $, 
\CheckRmv{
  \begin{equation*}
    r_{k}^{{\rm low}}=\left\{\begin{array}{ll}
      -\infty,& k=1, \\
      \left(k-2^{b-1}-1\right) \Delta, &k=2, \ldots, 2^{b},  
      \end{array}\right.
  \end{equation*}
}
and
\CheckRmv{
  \begin{equation*}
    r_{k}^{{\rm up}}=\left\{\begin{array}{ll}
      \left(k-2^{b-1}\right) \Delta, &k=1, \ldots, 2^{b}-1,  \\
      +\infty,& k=2^b,
      \end{array}\right.
  \end{equation*}
}
where $\Delta$ is the step size.
For the selection of $\Delta$, we utilize the received power-dependent step size  \cite{fesl2024channel}, 
\CheckRmv{
  \begin{equation}
    \Delta = \sqrt{P_y/2}\Delta_{\rm b},
  \end{equation}
}
where $P_y=\mathbb{E}[\|\mathbf{y}\|_2^2]$ is the received power, and $\Delta_{\rm b}$ is the step size optimized for zero-mean unit variance Gaussian inputs \cite{max1960quantizing}. In practical implementation, this variable step size can be realized by automatic gain control.

When low-resolution ADCs are employed at the receiver, severe nonlinear effects exacerbate the difficulty of the channel estimation task.  
Therefore, the significance of prior information becomes more prominent in this ill-posed inverse problem.
That being said, the pre-trained DM seamlessly fits this task as it exclusively models the channel prior distribution instead of learning the mapping from measurements to channel estimates.
Hence, the only thing to be modified compared to the infinite resolution case is the noise-perturbed likelihood for quantized measurements $\bar{\mathbf{y}}$ in \eqref{eq:model_quan}.

Building upon the preceding uninformative prior assumption and integrating \eqref{eq:ht_to_y} into \eqref{eq:model_quan}, we have
\CheckRmv{
  \begin{equation}
    \bar{\mathbf{y}} = Q(\frac{1}{\sqrt{\bar{\alpha}_t}}\mathbf{Ah}_t
    +\tilde{\mathbf{n}}),
  \end{equation}
}
where $\tilde{\mathbf{n}}\sim \mathcal{N}(\mathbf{0}, \mathbf{C}_t)$ with $\mathbf{C}_t=\frac{1-\bar{\alpha}_{t}}{\bar{\alpha}_{t}}\mathbf{A}\mathbf{A}^T+\sigma_{n}^{2}\mathbf{I}$.
Denoting $\boldsymbol{z}_t \triangleq \frac{\mathbf{A}\mathbf{h}_t}{\sqrt{\bar{\alpha}_t}}$, the noise-perturbed likelihood can be revised as \cite{meng2022quantized}
\CheckRmv{
  \begin{align}
    p_t&(\bar{\mathbf{y}}|\mathbf{h}_t) = \nonumber\\
    &\int \prod_{m=1}^M \mathbb{I}\big(z_m+\tilde{n}_m \in Q^{-1}(\bar{y}_m)\big)\cdot\mathcal{N}(\tilde{\mathbf{n}};\mathbf{0},\mathbf{C}_t){\rm d}\tilde{\mathbf{n}},
    \label{eq:pseudo_likelihood_quan}
  \end{align}
}
where $z_m$, $\tilde{n}_m$, and $\bar{y}_m$ denote the $m$-th element of $\boldsymbol{z}_t$, $\tilde{\mathbf{n}}_t$, and $\bar{\mathbf{y}}$, respectively.  The notation $Q^{-1}(\bar{y}_m)$ represents the quantization interval associated with the quantized output $\bar{y}_m$.

It is difficult to exactly compute \eqref{eq:pseudo_likelihood_quan}  since $\mathbf{C}_t$ is generally a non-diagonal matrix 
\cite{meng2022quantized}, resulting in a computationally expensive high-dimensional integral.
To simplify, and motivated by \cite{meng2022quantized}, we further assume that $\mathbf{A}$ is row-orthogonal such that $\mathbf{AA}^{T}$ reduces to a diagonal matrix.  
By combining this assumption with the previous uninformative prior assumption, we establish the following proposition for a closed-form approximation of $\nabla_{\mathbf{h}_t} \log p_t(\bar{\mathbf{y}}|\mathbf{h}_t)$. The detailed derivation is left in Appendix~\ref{appendix:quan}.

\begin{proposition}
  \label{quantized_likelihood}
   Assuming an uninformative prior, i.e., $p(\mathbf{h}_0|\mathbf{h}_t) \propto p(\mathbf{h}_t|\mathbf{h}_0)p_0(\mathbf{h}_0)\approx p(\mathbf{h}_t|\mathbf{h}_0)$, and a row-orthogonal $\mathbf{A}$, the noise-perturbed likelihood score for $\bar{\mathbf{y}}$ can be derived in closed form as
   \CheckRmv{
    \begin{equation}
      \nabla_{\mathbf{h}_t} \log p_t(\bar{\mathbf{y}}|\mathbf{h}_t) = {\frac{1}{\sqrt{\bar{\alpha}}_t}} \mathbf{A}^T\mathbf{g},
      \label{eq:pseudo_likelihood_quan_score}
    \end{equation}
   }
   where $\mathbf{g} = [g_{1}, g_{2},\ldots,g_{M}]^T\in \mathbb{R}^{M\times 1}$, the $m$-th ($m\in[M]$) element given by
   \CheckRmv{
    \begin{equation}
      g_{m}=\frac{\exp \Big(-\frac{(\tilde{y}_{m}^{\rm{low}})^{2}}{2}\Big)-\exp \Big(-\frac{(\tilde{y}_{m}^{\rm{up}})^{2}}{2}\Big)}
      {\sqrt{2\pi}\tilde{\sigma}_m \left(\Phi(\tilde{y}_{m}^{\rm{up}}) - \Phi(\tilde{y}_{m}^{\rm{low}}) \right)}, 
      \label{eq:g_real}
    \end{equation}
   }
   where 
   \CheckRmv{
    \begin{equation}
      \tilde{y}_{m}^{\rm{up}} = \frac{\bar{y}_{m}^{\rm{up}}- z_m}{\tilde{\sigma}_m},\;\tilde{y}_{m}^{\rm{low}} = \frac{\bar{y}_{m}^{\rm{low}}- z_m}{\tilde{\sigma}_m}, 
      \label{eq:modified_threshold}
    \end{equation}
   }
   $\Phi(u)=\frac{1}{\sqrt{2 \pi}} \int_{-\infty}^{u} e^{-\frac{z^{2}}{2}} {\rm d} z$ denotes the cumulative distribution function of the standard Gaussian distribution,
   and $\tilde{\sigma}_m^2 = \frac{1-\bar{\alpha}_t}{\bar{\alpha}_t}\|\mathbf{a}_m^T\|_2^2+\sigma_n^2$ with $\mathbf{a}_m^T \in \mathbb{R}^{1\times N}$ denoting the $m$-th row of $\mathbf{A}$. In \eqref{eq:modified_threshold}, $\bar{y}_{m}^{\rm low}$ and $\bar{y}_{m}^{\rm up}$ denote the lower and upper thresholds with respect to the codeword $\bar{y}_{m}$, respectively.
\end{proposition}

Based on Proposition \ref{quantized_likelihood}, we can adapt \algref{alg:dmce} to  low-resolution quantization by substituting Line 5 of the algorithm with the modified noise-perturbed likelihood score in \eqref{eq:pseudo_likelihood_quan_score}.

\begin{remark}
  To gain insights, the row-orthogonal assumption of $\mathbf{A}$ renders the  covariance of the effective noise $\tilde{\mathbf{n}}$ diagonal. In other words, the elements of $\tilde{\mathbf{n}}$ would be independent of each other. Therefore, the high-dimensional integral involved in \eqref{eq:pseudo_likelihood_quan} can be decoupled, leading to significant simplification of computation. 
  For practical considerations, this assumption can be met through the use of orthogonal pilots.
\end{remark}

\subsection{Learning from Noisy Channel Realizations} \label{sec:sure-dm} 
In practical air interfaces, acquiring a large dataset comprising ground truth channel samples for training DL-based channel estimators is expensive or impossible. 
A potential approach to addressing this challenge is to enable learning the models solely using noisy channel realizations as the training data \cite{fesl2024channel,doshi2022over,aali2023solving}. 
We investigate the adaptation of the proposed approach in this subsection.
Specifically, we consider an LMMSE estimator for acquiring the training channel dataset from noisy pilot measurements. 
The acquired training sample is thus statistically equivalent to an AWGN observation of the true channel $\mathbf{h}$, i.e.,  
\CheckRmv{
  \begin{equation}
    \tilde{\mathbf{h}} = \mathbf{h} + \mathbf{w}, 
    \label{eq:awgn}
  \end{equation}
}
where $\mathbf{w}\sim \mathcal{N}(\mathbf{0},\sigma_w^2\mathbf{I})$ is the estimation error with $\sigma_w^2$ denoting the error variance \cite{weberImperfectChannelstateInformation2006}.
We further normalize $\tilde{\mathbf{h}}$ to have unit power to align it with the noise model of the DM in \eqref{eq:h0_to_ht}, 
\CheckRmv{
  \begin{equation}
    \bar{\mathbf{h}} = \sqrt{\bar{\alpha}_{t_w}} \mathbf{h} + \sqrt{1-\bar{\alpha}_{t_w}}\boldsymbol{\epsilon},
    \label{eq:vp-sde}
  \end{equation}
}
where $\bar{\mathbf{h}}=\sqrt{\bar{\alpha}_{t_w}} \tilde{\mathbf{h}}$, $\bar{\alpha}_{t_w}=\frac{1}{1+\sigma_w^2}$, and 
\CheckRmv{
  \begin{equation}
    t_w = \underset{t\in [T]}{\arg \min} \;|\bar{\alpha}_t - \bar{\alpha}_{t_w} |.
  \end{equation}
}

Inspired by \cite{aali2023solving}, we develop a training strategy for the DM when only a noisy channel dataset $\bar{\mathcal{H}} = \{\bar{\mathbf{h}}^{(i)}\}$ is available by coupling the SURE denoising into the training of DM. To begin, we provide a brief introduction to the SURE denoising.

The SURE is a classical solution for learning a denoiser to reconstruct data from noisy observations without access to ground truth data \cite{stein1981estimation}. 
It offers an unbiased estimate of the MSE loss 
that necessitates the ground truth $\mathbf{h}$.  
The SURE loss for \eqref{eq:vp-sde} is given by \cite{eldar2008generalized} 
\CheckRmv{
  \begin{align}
    \mathcal{L}&_{\mathrm{SURE},\sigma_w}(\boldsymbol{\theta}) \nonumber \\
    &=\mathbb{E}_{\bar{\mathbf{h}}}\left[\left\| f_{\boldsymbol{\theta}}(\bar{\mathbf{h}})-\frac{1}{\sqrt{\bar{\alpha}_{t_w}}}\bar{\mathbf{h}}\right\|_2^2+2\sigma_w^2\cdot\mathrm{div}_{\bar{\mathbf{h}}}(f_{\boldsymbol{\theta}}(\bar{\mathbf{h}}))\right],
    \label{eq:sure_loss}
  \end{align}
}
where $f_{\boldsymbol{\theta}}(\cdot)$ is the denoiser that takes the noisy $\bar{\mathbf{h}}$ as input and outputs an estimate of $\mathbf{h}$, and $\mathrm{div}_{\bar{\mathbf{h}}}(\cdot)$ denotes the divergence defined as the sum of partial derivates with respect to each element of $\bar{\mathbf{h}}$
given by 
\CheckRmv{
  \begin{equation}
    \mathrm{div}_{\bar{\mathbf{h}}}(f_{\boldsymbol{\theta}}(\bar{\mathbf{h}})) = \sum_{n=1}^{N} \frac{\partial [f_{\boldsymbol{\theta}}(\bar{\mathbf{h}})]_n}{\partial \bar{h}_n}.
  \end{equation}
}
To avoid the high-complexity $N$-times partial derivates involved in this definition, we follow the well-established practice \cite{ramaniMonteCarloSureBlackBox2008} to simplify computation by employing a Monte Carlo approximation as follows
\CheckRmv{
  \begin{equation}
    \mathrm{div}_{\bar{\mathbf{h}}}(f_{\boldsymbol{\theta}}(\bar{\mathbf{h}}))\approx\mathbf{v}^T\left(\frac{f_{\boldsymbol{\theta}}(\bar{\mathbf{h}}+\varepsilon\mathbf{v})-f_{\boldsymbol{\theta}}(\bar{\mathbf{h}})}{\varepsilon}\right),
  \end{equation}
} 
where $\mathbf{v}\sim \mathcal{N}(\mathbf{0},\mathbf{I})$, and $\varepsilon$ is a small positive number, which is set to $10^{-5}$ in experiments according to \cite{nguyen2020hyperspectral}.

\CheckRmv{
  \begin{figure}[t]
    \centering
    \includegraphics[width=3.45in]{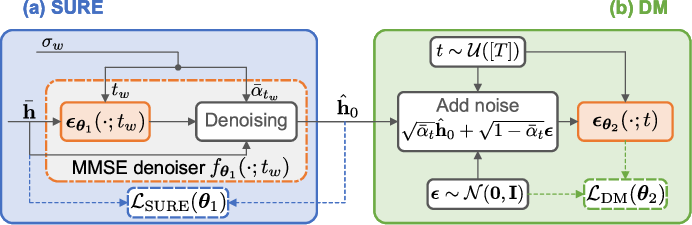}
    \caption{Block diagram of the SURE-DM's training flow.}
    \label{fig:dm_sure}
  \end{figure}
}

Collectively, the proposed method adds noise to the MMSE-denoised channel samples obtained by SURE denoising and learns the corresponding generative prior based on DM's training objective.
The SURE denoising and DM's training can be unified by 
Tweedie's formula \cite{efron2011tweedie}, enabling both parts to share the same network structure $\boldsymbol{\epsilon}_{\boldsymbol{\theta}}$ while using different parameters, ${\boldsymbol{\theta}}_1$ and ${\boldsymbol{\theta}}_2$. {We adopt separate weights for the SURE and DM due to the fundamentally different denoising objectives of the two components, with SURE targeting a single noise level and DM adapting to multiple noise levels.} 
The block diagram of the proposed training scheme, termed SURE-DM, is presented in \figref{fig:dm_sure}.

Specifically, for the noise model in \eqref{eq:vp-sde}, Tweedie's formula \cite{efron2011tweedie} introduces a MMSE denoiser,
\CheckRmv{
  \begin{equation}
    \mathbb{E}[\mathbf{h}|\bar{\mathbf{h}}]=\frac{1}{\sqrt{\bar{\alpha}_{t_w}}}\big(\bar{\mathbf{h}}+(1-\bar{\alpha}_{t_w})\nabla_{\bar{\mathbf{h}}}\log p(\bar{\mathbf{h}})\big).
  \end{equation}
}
We can replace $\nabla_{\bar{\mathbf{h}}}\log p(\bar{\mathbf{h}})$ with the score estimate on the right-hand side (RHS) of \eqref{eq:score_dm} and approximate the MMSE denoiser as  
\CheckRmv{
  \begin{equation}
    f_{\boldsymbol{\theta}_1}(\bar{\mathbf{h}},{t_w})=\frac{1}{\sqrt{\bar{\alpha}_{t_w}}}\big(\bar{\mathbf{h}}-\sqrt{1-\bar{\alpha}_{t_w}}\boldsymbol{\epsilon}_{\boldsymbol{\theta}_1}(\bar{\mathbf{h}},{t_w})\big).
    \label{eq:mmse_denoiser}
  \end{equation}
}
This approximate MMSE denoiser can be learned using the SURE loss in \eqref{eq:sure_loss}. 
After the convergence of the training for the MMSE denoiser, the parameters $\boldsymbol{\theta}_1$ are fixed, and the DM's denoising network $\boldsymbol{\epsilon}_{\boldsymbol{\theta}_2}$ can be trained based on the denoised samples $\hat{\mathbf{h}}_0=f_{\boldsymbol{\theta}_1}(\bar{\mathbf{h}},{t_w})$. This training process employs the loss function from \eqref{eq:dm_loss}, where the ground truth $\mathbf{h}_0$ is replaced by the denoised samples $\hat{\mathbf{h}}_0$.

We note that although this strategy differs from learning directly on ground truth samples, the SURE-denoised samples serve as an effective alternative, as verified by simulation results. Therefore, this strategy offers an approach to learning a DM for wireless channels without the need for a significant amount of clean channel data, which can be challenging to acquire in practice.

\section{Numerical Results}
This section presents numerical results to evaluate the proposed DM-based channel estimator. We initiate by introducing experimental details including datasets and training setups. Then, we investigate the channel estimation performance of the proposed method in both full- and low-resolution receivers. Next, we conduct experiments to validate the effectiveness of the proposed SURE-DM scheme. Finally, we provide a detailed complexity analysis.

\subsection{Experimental Details} \label{sec:simu_setup}
We adopt the QuaDRiGa toolbox \cite{jaeckelQuaDRiGaQuasiDeterministic} as the channel simulator to generate training, testing, and validation data. 
We consider an urban macro-cell line-of-sight (LOS) scenario according to the 3rd generation partnership project (3GPP) 38.901 technical report \cite{3gpp38901}. 
The carrier frequency is set to 40 GHz. ULAs with ``3GPP-3D'' antennas featuring half-wavelength spacing are utilized at both the base station (BS) and the multi-antenna user terminal (UT). Downlink transmission is assumed, where the BS's antenna array is placed at a height of 25 m. The UT randomly resides in a $120^{\circ}$ cell sector with a radius of 500 m and a guard distance of 35 m to the BS. The number of antennas at the BS and UT is set as $\nt=64$ and $\nr=16$, respectively.
{A total of $D=100,000$ channel samples are generated as the training dataset, each with different UT locations to enhance diversity.} Two distinct setups are considered, i.e., noiseless and noisy datasets. For the noiseless case (Sections \ref{sec:simu_full} and \ref{sec:simu_low}), a simulated dataset $\mathcal{H}= \{\mathbf{h}^{(i)}\}_{i=1}^D$ containing ground truth samples generated via QuaDRiGa 
is employed; for the noisy case (Section \ref{sec:dm_sure}), the simulated dataset $\mathcal{H}$ serves as a benchmark for deriving 
normalized LMMSE estimates as given in \eqref{eq:vp-sde}, 
constituting a noisy dataset $\bar{\mathcal{H}}=\{\bar{\mathbf{h}}^{(i)}\}_{i=1}^D$.
Furthermore, a validation dataset containing 100 channel realizations is utilized for hyperparameter tuning.

During training, the number of epochs is set to 500. The batch size is configured as 128. The Adam optimizer with the learning rate fixed as $10^{-4}$ is utilized.
For the noise schedule $\{\beta_t\}_{t=1}^T$ of DM, we employ the linear schedule\footnote{{We also tried other advanced noise schedules, including the quadratic \cite{song2020improved} and exponential schedules \cite{karras2022elucidating}, but did not observe performance enhancement.}} prescribed in \cite[Table 1]{fesl2024asymptotic} for different numbers of time steps $T$. 
For the backbone CNN of the denoising network, the maximum channel size is set as $S_{\max}=64$, leading to the channel sizes of different convolutional layers as shown in \figref{fig:network}. Details of the network are illustrated in Appendix~\ref{appendix:net}.
The gradient scale $s$ of the proposed approach is selected by grid search using the validation dataset. 
During testing, we generate 100 channel realizations with UT locations that differ from the training dataset to assess the estimation normalized MSE (NMSE) performance defined as
\CheckRmv{
  \begin{equation}
    \text{NMSE} = \mathbb{E}\frac{\|\hat{\mathbf{h}} - \mathbf{h}\|_2^2}{\|\mathbf{h}\|_2^2}.
  \end{equation}
}
{The channel samples are normalized by utilizing the average channel power calculated from the training dataset across all data samples and channel entries, ensuring consistency with the definition of SNR adopted in the system model.}
The pilot symbols are randomly drawn from the QPSK lattice unless noted otherwise.

The following baseline methods are compared: 
\begin{itemize}
  \item \textbf{LMMSE}: Linear estimator using the sample covariance computed via all training samples, $\mathbf{C}_h = \frac{1}{D}\sum_{i=1}^D \mathbf{h}^{(i)}(\mathbf{h}^{(i)})^{T}$, and computing the estimate as $\hat{\mathbf{h}} = \mathbf{C}_h \mathbf{A}^{T}(\mathbf{A}^{T}\mathbf{C}_h\mathbf{A} + \sigma_n^2\mathbf{I})^{-1}\mathbf{y}$ \cite{kay1993fundamentals}.
  \item \textbf{Bussgang LMMSE (BLMMSE)}: The linear estimator introduced in\cite{li2017channel} that adapts to low-resolution measurements based on Bussgang decomposition \cite{bussgang1952crosscorrelation}. 
  \item \textbf{LASSO}: A standard $l_1$-norm regularization-based CS channel recovery method within the angular domain \cite{schniterChannelEstimationPrecoder2014}. 
  \item \textbf{EM-GM-AMP}: The AMP-based iterative CS scheme from \cite{vilaExpectationMaximizationGaussianMixtureApproximate2013}, which adopts a Gaussian mixture prior for angular domain channels. 
  \item \textbf{LDAMP}: A supervised DL-based channel estimator by unfolding the AMP method and incorporating a denoising CNN \cite{heDeepLearningBasedChannel2018}. 
  We use the implementation from the publicly available code.\footnote{\url{https://github.com/hehengtao/LDAMP_based-Channel-estimation}}
  \item \textbf{Variational autoencoder (VAE)}: The VAE-based channel estimator  that models the channel distribution as conditionally Gaussian. During inference, the estimator parameterizes the channel covariance for each received measurement, based on which individual LMMSE filters are employed to derive the final estimates \cite{fesl2024channel,baurLeveragingVariationalAutoencoders2023}.
  We adopt the publicly accessible implementation.\footnote{\url{https://github.com/benediktfesl/Quantized_Channel_Estimation}} 
  \item \textbf{SGM}: The SGM-based estimator from \cite{arvinteMIMOChannelEstimation2022}, which learns a score-based generative prior and performs estimation using annealed Langevin dynamics. We utilize the reference code repository\footnote{\url{https://github.com/utcsilab/score-based-channels}} for implementation and adopt the deepest model in \cite{arvinteMIMOChannelEstimation2022} for maximizing performance. 
\end{itemize}
Hyperparameters of these baselines are meticulously tuned based on the validation dataset. {Note that for the LDAMP and VAE methods, a distinct model is trained for each value of SNR, pilot density $\alpha$, and quantization bit count to maximize their performance due to their supervised learning nature. This results in substantial memory overhead compared to the proposed approach, which adopts a single model that is agnostic to these variables, enabling it to efficiently handle various configurations without the need for retraining.} {The GAN-based channel estimator from \cite{baleviHighDimensionalChannel2021} is not selected as a baseline due to its performance saturation at high SNRs, as demonstrated in \cite{arvinteMIMOChannelEstimation2022}.}

\subsection{Full-Resolution Channel Estimation} \label{sec:simu_full}

\CheckRmv{
  \begin{figure}[t]
    \centering
    \includegraphics[width=3.2in]{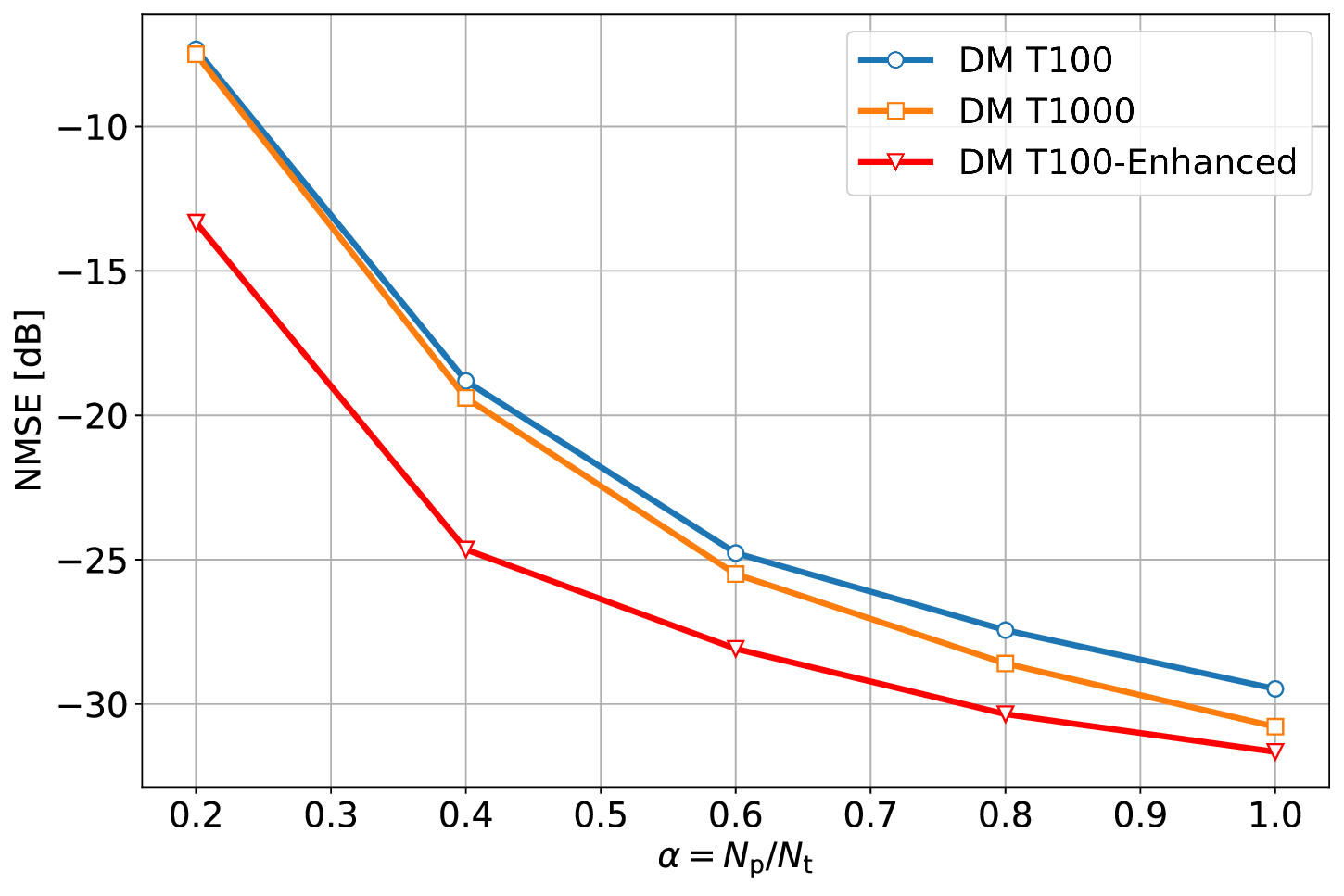}
    \caption{NMSE with respect to the pilot density of the proposed DM-based channel estimator across various configurations (\SNR{=}{30}).}
    \label{fig:dm_diff_alpha}
  \end{figure}
}

\CheckRmv{
  \begin{figure}[t]
    \centering
    \subfigure[{NMSE}]{
      \includegraphics[width=3.2in]{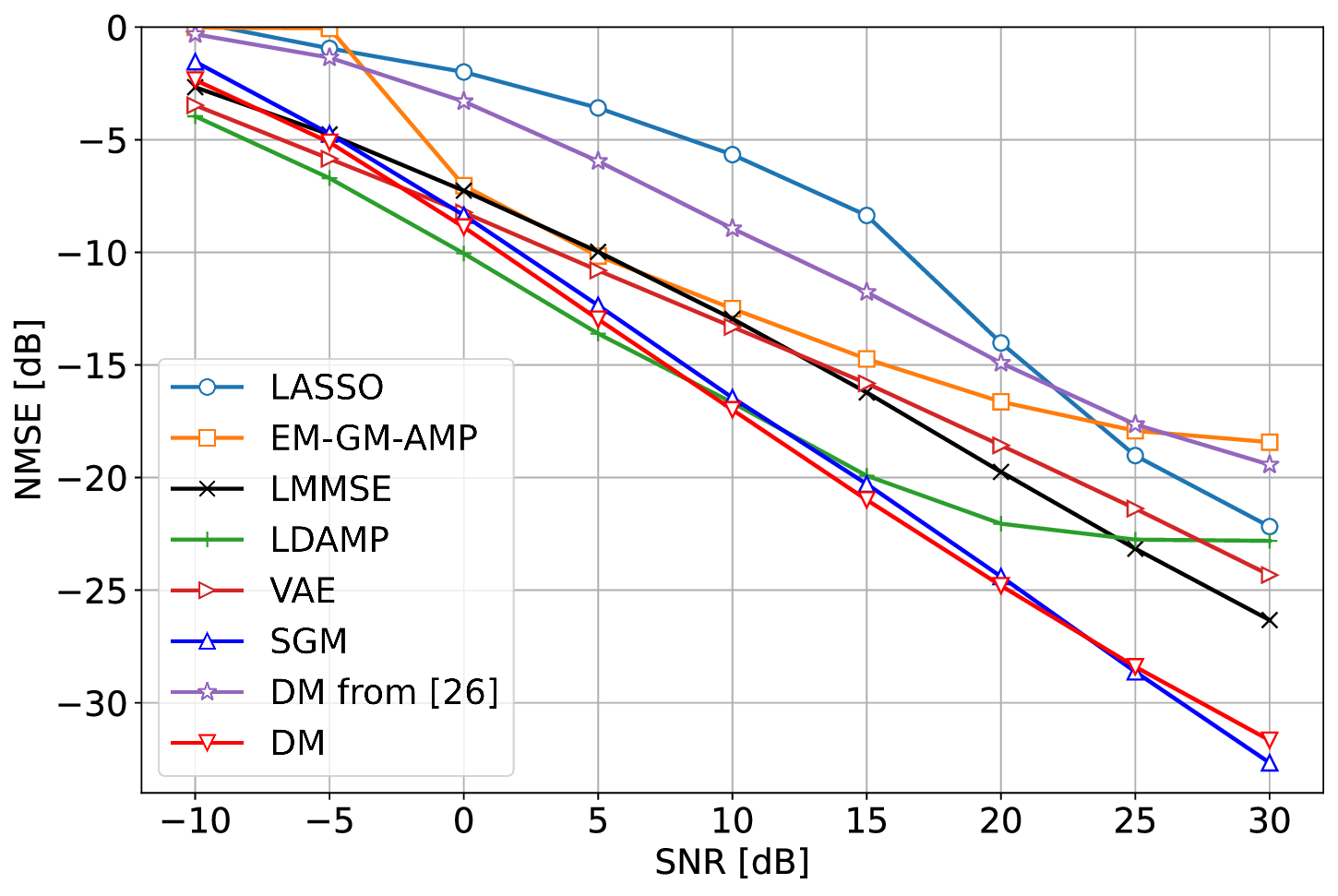}
      \label{fig:full_ce_alpha1_nmse}
    }
    \subfigure[{Latency}]{
      \includegraphics[width=3.2in]{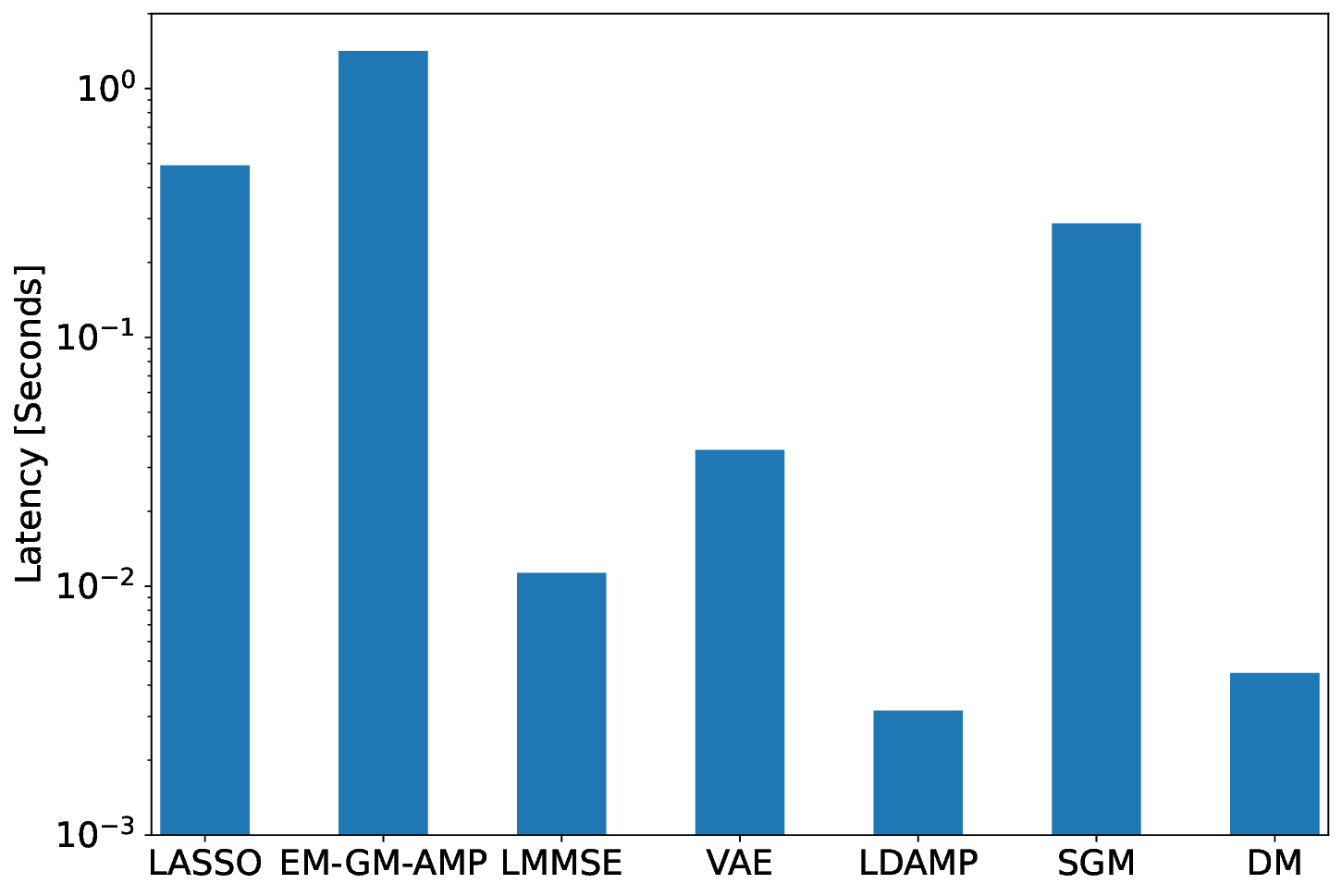}
      \label{fig:full_ce_alpha1_time}
    }
    \caption{{NMSE performance and estimation latency when $\alpha=1$.}}
    \label{fig:full_ce_alpha1}
  \end{figure}
}

We first evaluate the estimation performance of the proposed DM-based method in the full-resolution receiver. \figref{fig:dm_diff_alpha} presents the NMSE as a function of the pilot density $\alpha$ with the SNR fixed as 30 dB. 
We consider three distinct configurations for the proposed method. The former two are equipped with $T=100$ and $T=1000$, respectively. The last one, also equipped with $T=100$ time steps, utilizes an enhancing strategy to refine the estimate by iterating the update rule 3 rounds at each time index $t\in [0, T/2]$, i.e., the latter half of the reverse process. {This strategy, termed ``DM T100-Enhanced'', is motivated by the observation that there exists a refinement stage as $t$ approaches 0, during which using additional sampling iterations can significantly enhance the estimation performance \cite{yu2023freedom,song2023solving}.} As shown in \figref{fig:full_ce_alpha1_nmse}, the ``DM T100-Enhanced'' variant outperforms the vanilla versions with both $T=100$ and $T=1000$, particularly when $\alpha$ is small. Similar phenomena are noted in the low SNR regime, showcasing the superiority of this variant. Therefore, this variant is maintained for the remaining experiments.
Moreover, the figure shows that the NMSE improvement within the range $\alpha \in (0.4, 1]$ is relatively gradual compared to the rapid decrease in estimation error within the range $\alpha\in[0.2, 0.4]$. This result reveals that the prior incorporated in the DM can provide sufficient information for approaching the recovery performance achieved in a fully determined inverse problem.

\figref{fig:full_ce_alpha1} compares the NMSE and latency of the proposed method with baselines when the pilot density is $\alpha=1$. {As shown in \figref{fig:full_ce_alpha1_nmse}, the proposed DM-based approach outperforms all the compared schemes by a wide margin in estimation accuracy except for the high-complexity SGM.
Due to the absence of true covariance information, the LMMSE estimator only uses an estimated sample covariance, hence lagging behind the DM that can capture deep-level propagation features.
In addition, we compare the proposed approach with the DM-based channel estimator introduced in \cite{fesl2024diffusion}, which fails to achieve the same performance as the proposed approach. This disparity can be attributed to the fact that the theoretical framework and algorithm design presented in \cite{fesl2024diffusion} rely on the assumption of orthogonal pilot sequences and full pilot measurements. In contrast, the proposed method is not restricted by this assumption, enabling its applicability to scenarios with non-orthogonal pilot sequences, under-determined measurements (i.e., $\np<\nt$), and low-resolution ADCs. Given the limitations of the DM-based approach in \cite{fesl2024diffusion} within these contexts, we have opted not to include it as a baseline for comparison in the subsequent simulations.}
\figref{fig:full_ce_alpha1_time} presents the online estimation latency per channel realization across the compared methods. The latency is evaluated on a machine with an NVIDIA Tesla V100 GPU and an Intel Xeon E5-4627 CPU. The figure demonstrates that the proposed method achieves notable latency reduction compared to the SGM, making it one of the most efficient methods among those compared. This advantage can be attributed to the lightweight network architecture, substantially reduced time steps, and low computational overhead at each step. Considering the comparable performance to SGM as depicted in \figref{fig:full_ce_alpha1_nmse}, our method can be viewed as a favorable option for wireless channel estimation featuring real-time implementation. 

\CheckRmv{
  \begin{figure}[t]
    \centering
    \includegraphics[width=3.2in]{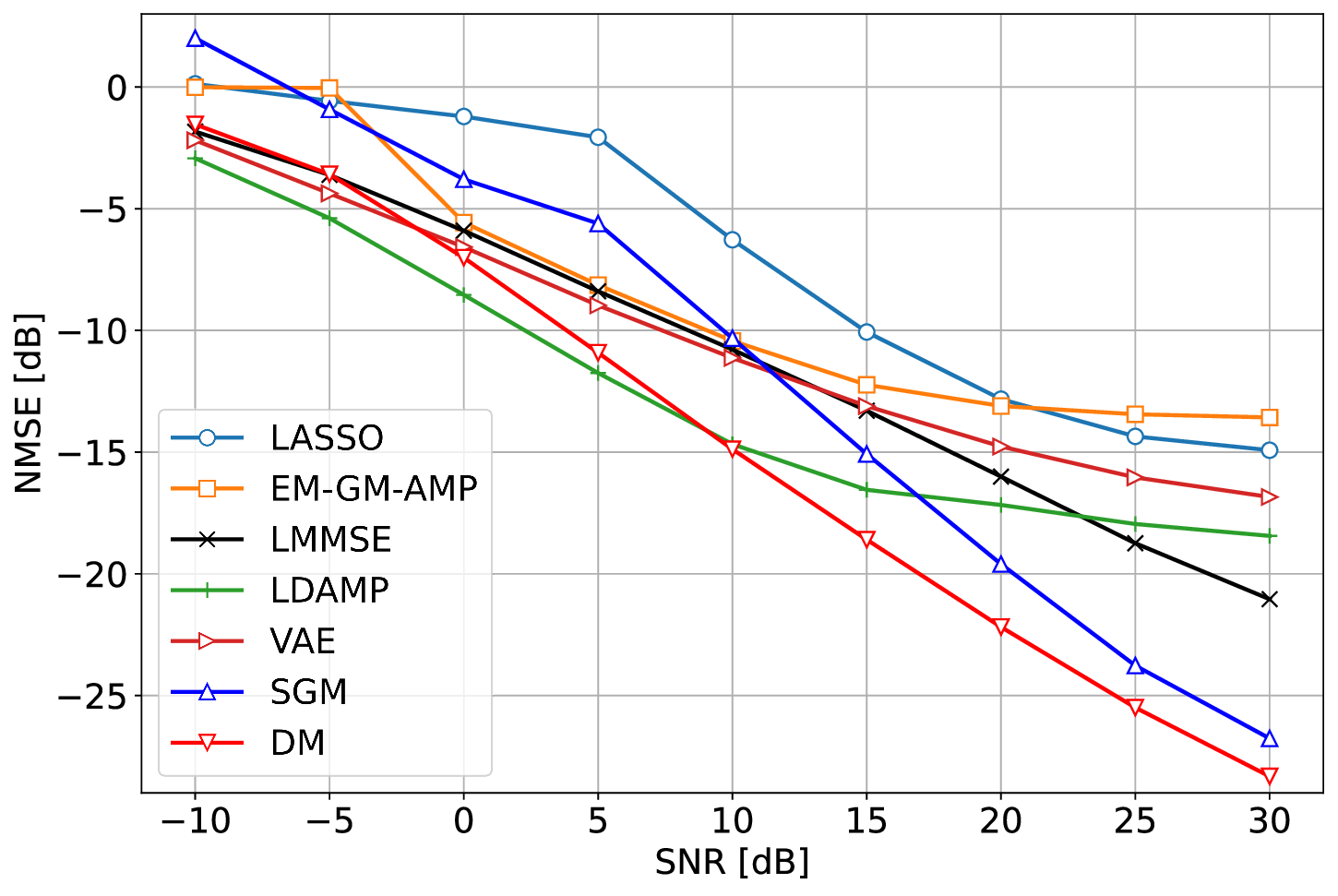}
    \caption{{NMSE performance when $\alpha=0.6$.}}
    \label{fig:full_ce}
  \end{figure}
}

\CheckRmv{
  \begin{figure}[t]
    \centering
    \includegraphics[width=3.45in]{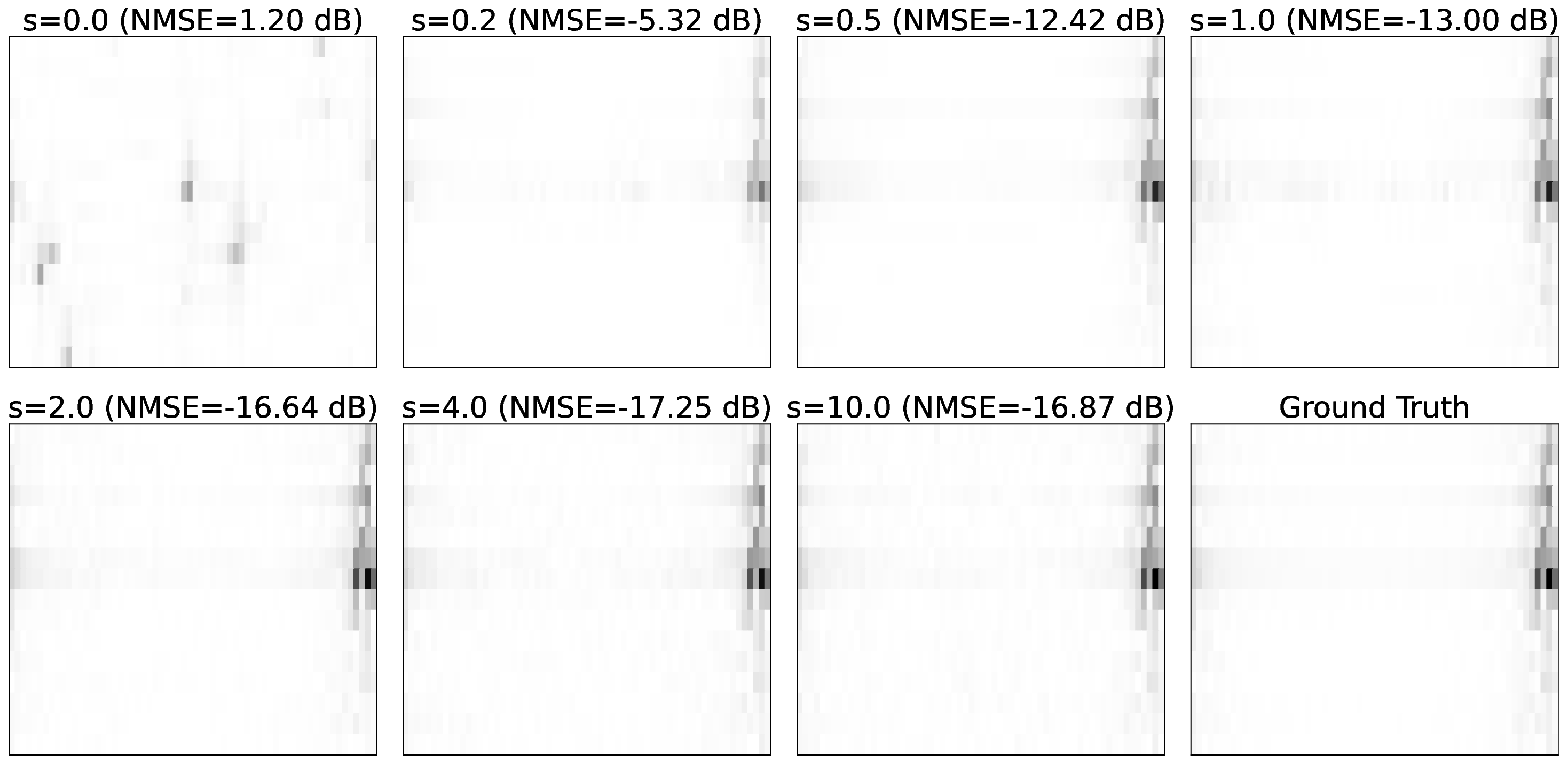}
    \caption{Recovered channel images and NMSE performance of the proposed method with different gradient scales. The SNR is 20 dB, and the pilot density is $\alpha=0.6$.} 
    \label{fig:full_ce_diff_scale}
  \end{figure}
}

We further investigate the impact of a limited number of pilots on performance comparisons. \figref{fig:full_ce} illustrates the NMSE when the pilot count is $\np=38$ ($\alpha=\np/\nt\approx 0.6$). As demonstrated in the figure, the DM-based method maintains its effectiveness under this limited pilot scenario owing to the powerful prior learned by the DM. The gains over baselines become even more pronounced. 

\CheckRmv{
  \begin{figure}[t]
    \centering
    \includegraphics[width=3.2in]{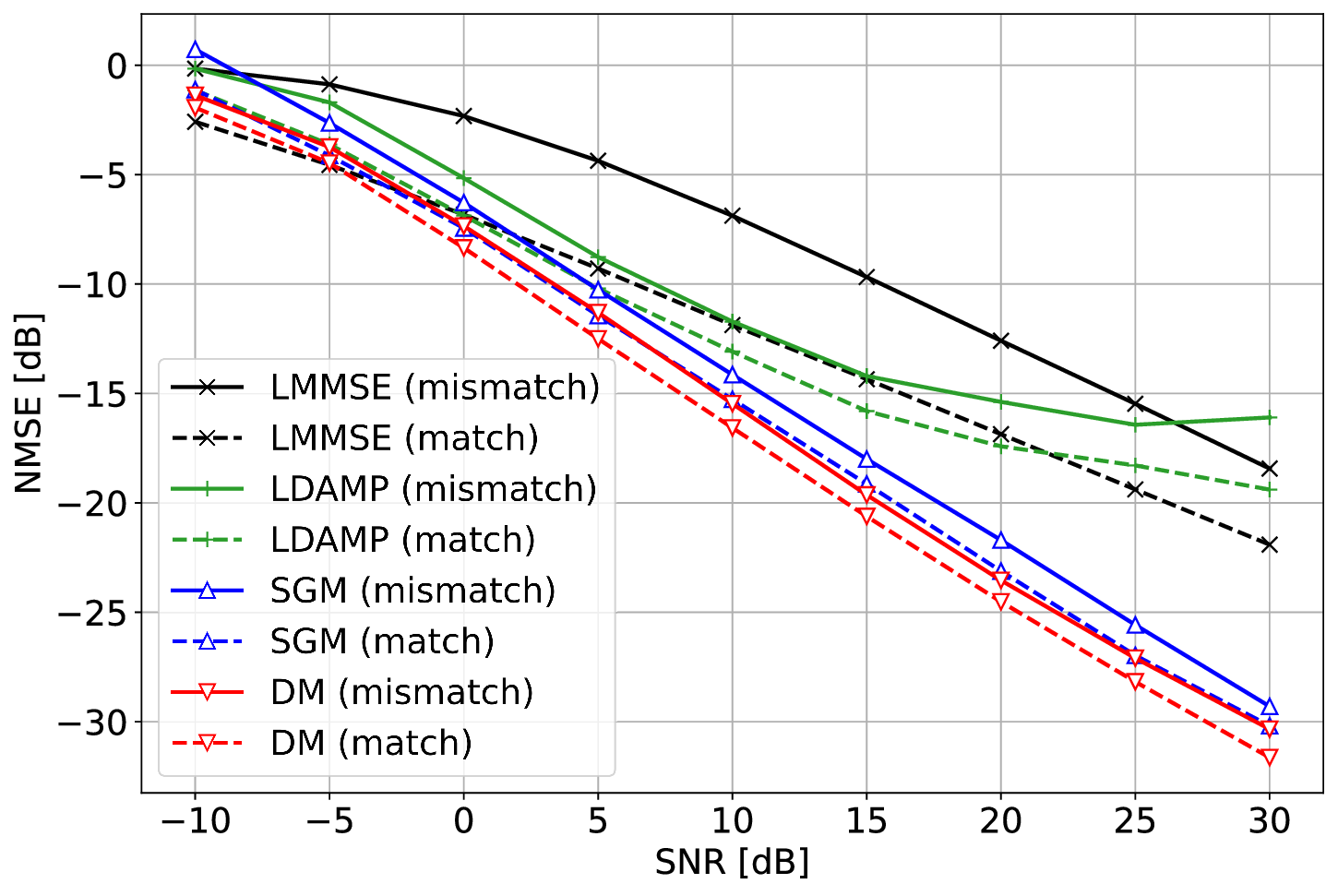}
    \caption{{Robustness of the proposed method under mismatch channel scenarios. We evaluate the performance in an NLOS scenario with $\alpha=1$.}}
    \label{fig:full_ce_robustness}
  \end{figure}
}

\figref{fig:full_ce_diff_scale} analyzes the performance of the proposed estimator using various choices of the gradient scale $s$ introduced in \algref{alg:dmce} by visualizing the recovered channel via grayscale plots. It is observed that the proposed method is empirically robust to the selection of this hyperparameter: when $s$ ranges from 1.0 to 10.0, the NMSE remains consistent and the recovered channels closely resemble the ground truth channel.

To verify the robustness of the proposed deep generative prior-based method, we consider a mismatched channel setting in \figref{fig:full_ce_robustness}. Specifically, we deploy the DM trained using LOS channel samples in a non-line-of-sight (NLOS) scenario (with other configurations remaining unchanged) for channel estimation without any retraining or adaptation.
We also present the NMSE of benchmark models trained with NLOS channel samples.
{As shown in the figure, the performance loss of the proposed method in mismatched setups is very small compared to that of the LMMSE and LDAMP, similar to the SGM baseline.}
This can be attributed to the powerful distribution coverage capability of generative DMs \cite{sohl2015deep}.   

\subsection{Low-Resolution Channel Estimation}  \label{sec:simu_low}
\CheckRmv{
  \begin{figure}[t]
    \centering
    \subfigure[NMSE]{
      \includegraphics[width=3.2in]{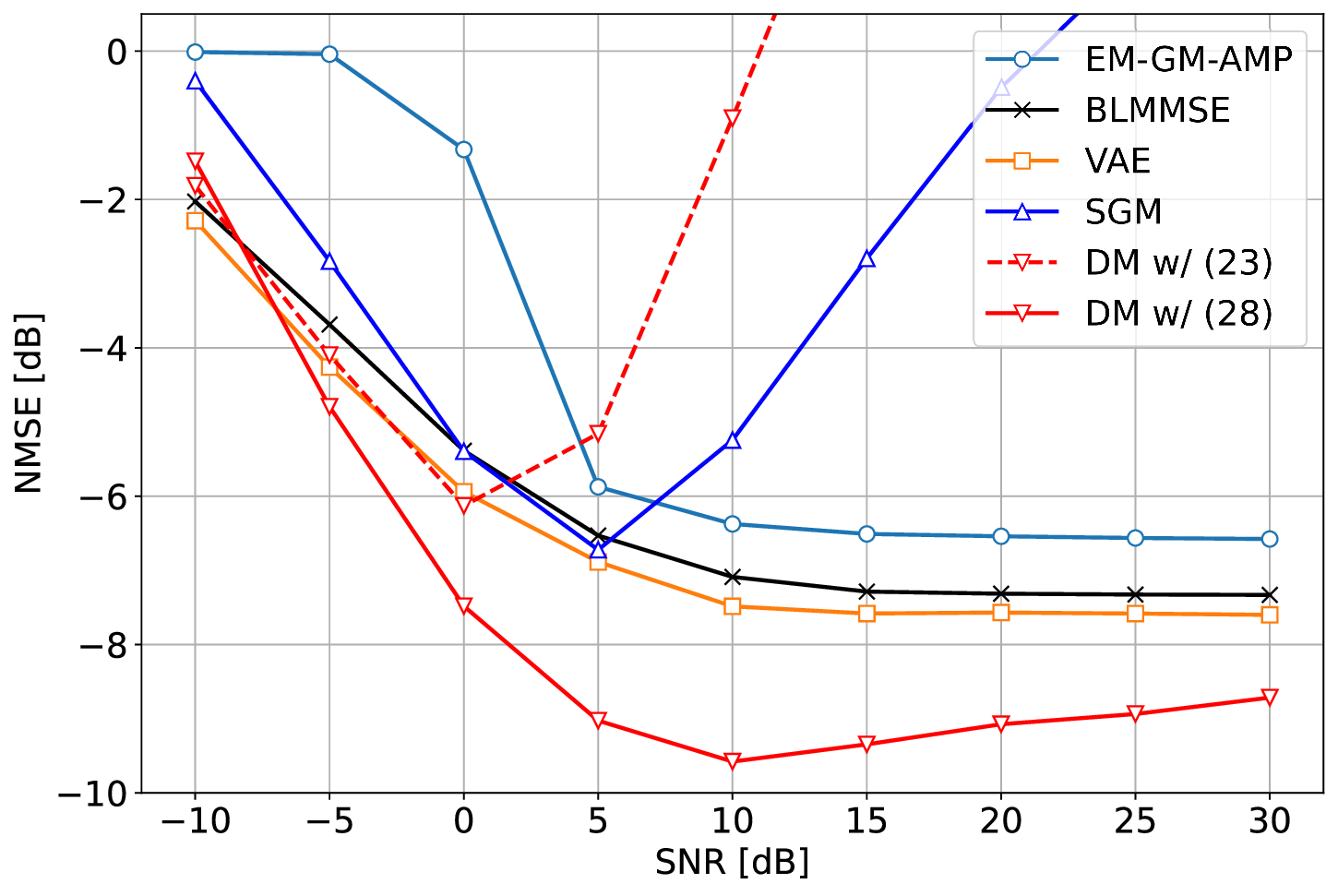}
      \label{fig:1bit_ce_alpha1_nmse}
    }
    \subfigure[Latency]{
      \includegraphics[width=3.2in]{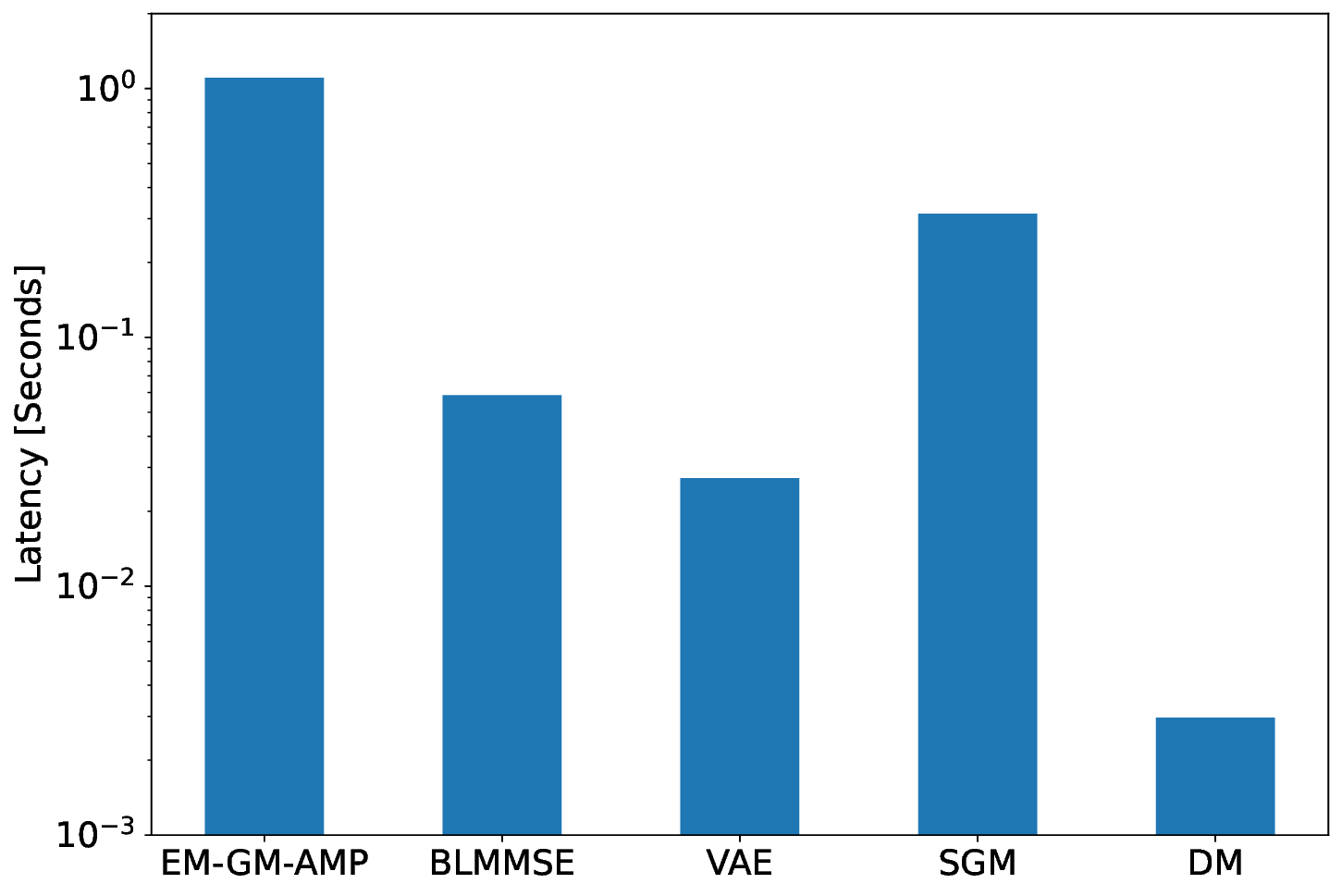}
      \label{fig:1bit_ce_alpha1_time}
    }
    \caption{NMSE performance and estimation latency under 1-bit ADCs when $\alpha=1$.}
    \label{fig:1bit_ce_alpha1}
  \end{figure}
}

In this subsection, we evaluate the channel estimation performance when the receiver is equipped with few-bit ADCs. 
Note that the pilot sequence (QPSK-based) is not specifically crafted to adhere to the row-orthogonal assumption of $\mathbf{A}$ as outlined in Proposition \ref{quantized_likelihood}. Nonetheless, results show the robustness of the update rule derived in Proposition \ref{quantized_likelihood} to the violation of this assumption.
We first consider the extreme 1-bit ADC scenario, where the quantization function becomes a sign function. The corresponding NMSE performance and estimation latency is shown in \figref{fig:1bit_ce_alpha1}.
\figref{fig:1bit_ce_alpha1_nmse} showcases that the DM-based channel estimator using the revised likelihood score for low-resolution ADCs, i.e., ``w/ \eqref{eq:pseudo_likelihood_quan_score}'', outperforms linear (BLMMSE), CS-based (EM-GM-AMP), and DL-based (VAE) estimators by more than 1 dB in NMSE. This figure also highlights the necessity of utilizing the revised score instead of the score for the full-resolution case in \eqref{eq:pseudo_likelihood_svd}.  
It is observed that both SGM and the proposed DM for full-resolution channel estimation exhibit severe degradation when \SNR{>}{5}, confirming the significance of the modification to tackle nonlinear effects caused by low-resolution ADCs. Therefore, we always utilize this modification for quantized channel estimation in the rest of the experiments. 

\figref{fig:1bit_ce_alpha1_time} further reveals the superiority of the proposed method in computation efficiency. The latency of DM is comparable to the low-complexity LS estimator and is significantly lower than other baselines, achieving a reduction of more than 10 folds. These results, combined with the performance gains over all baselines shown in \figref{fig:1bit_ce_alpha1_nmse}, underscore the practical feasibility and favorable properties of our method in massive MIMO channel estimation with few-bit ADCs. 

\CheckRmv{
  \begin{figure}[t]
    \centering
    \includegraphics[width=3.2in]{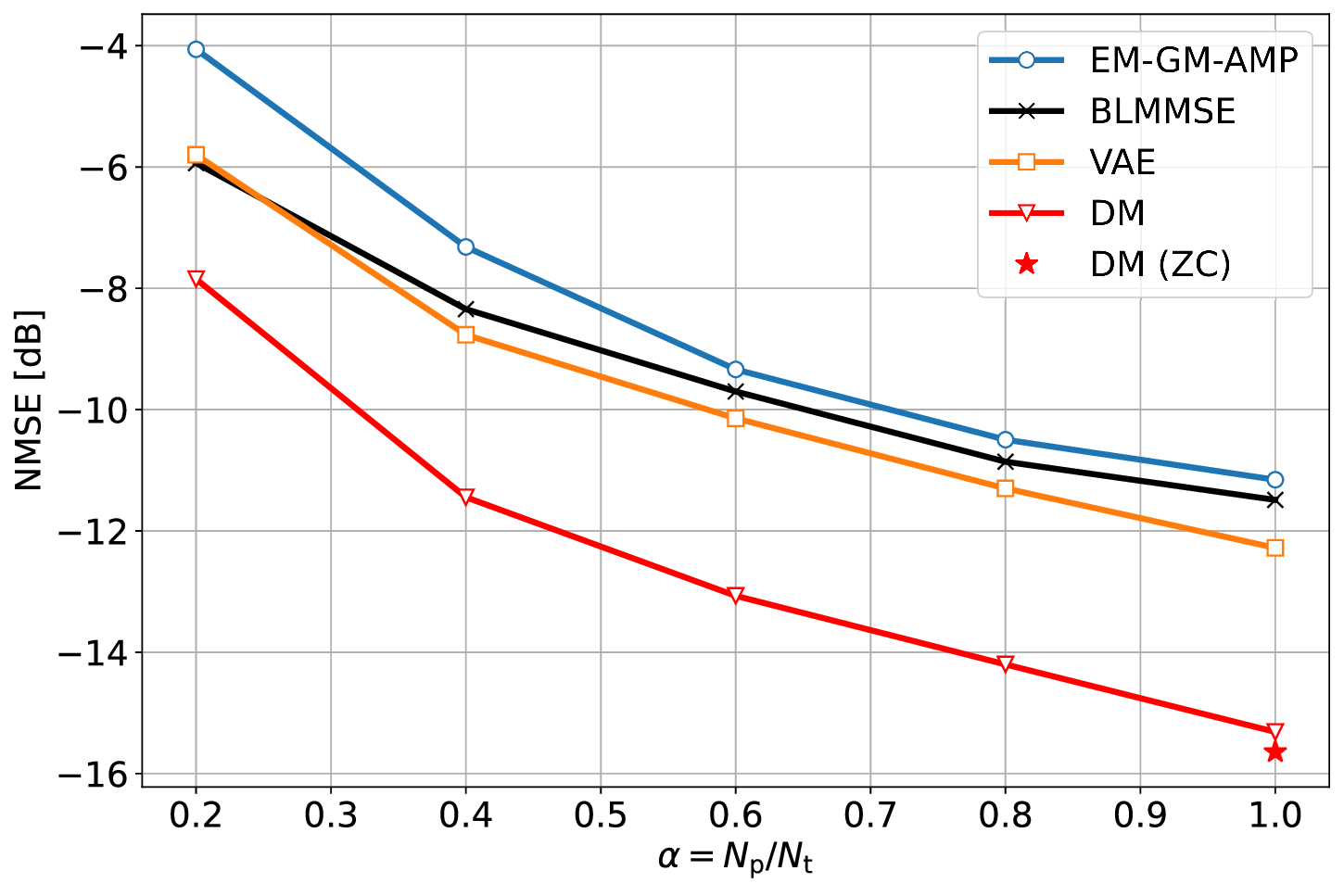}
    \caption{Performance with various pilot density $\alpha$ under 3-bit ADCs and \SNR{=}{10}.}
    \label{fig:3bit_ce_diff_alpha}
  \end{figure}
}

\CheckRmv{
  \begin{figure}[t]
    \centering
    \includegraphics[width=3.2in]{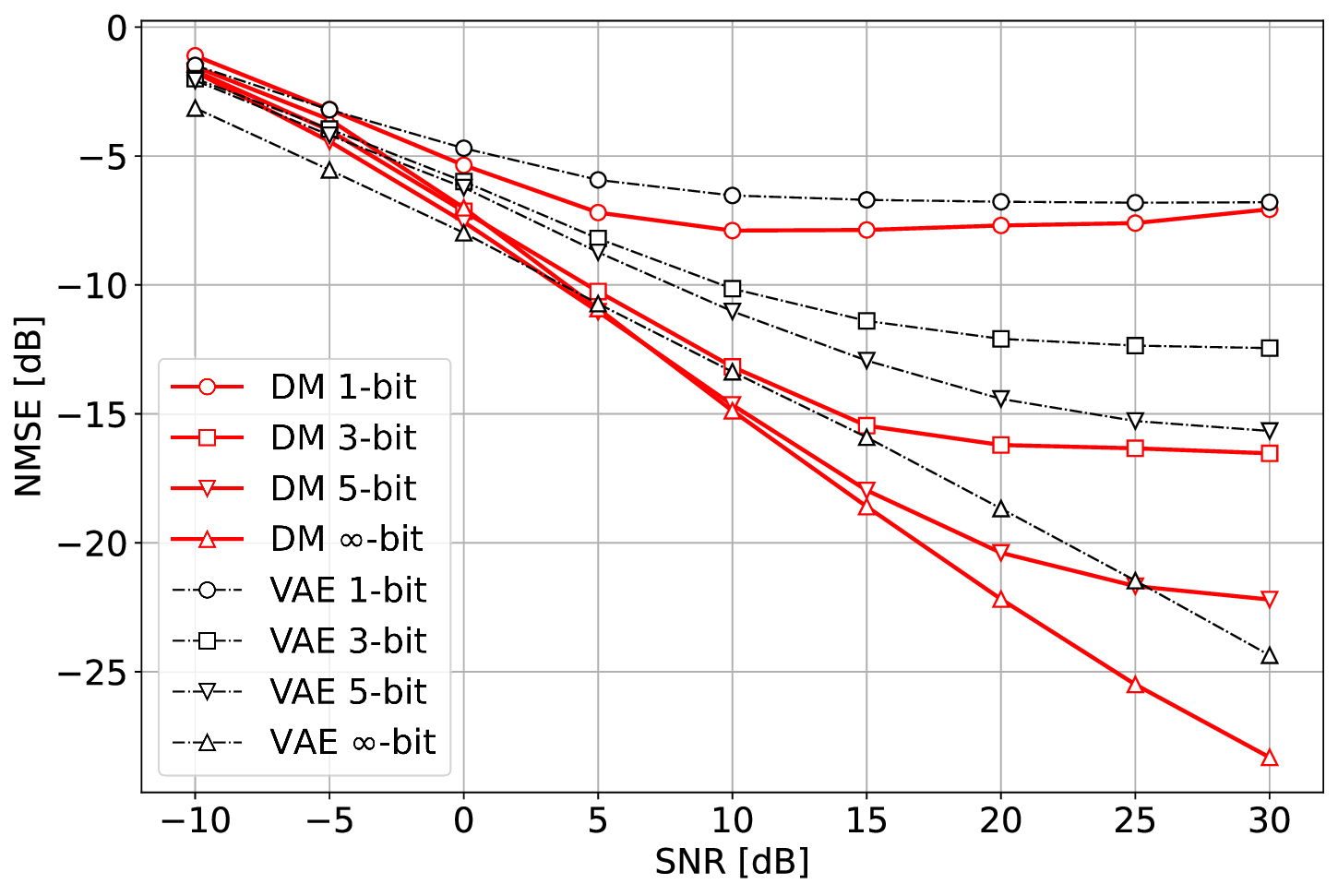}
    \caption{Performance with various ADC resolutions when $\alpha=0.6$.}
    \label{fig:lowbit_ce_diffbit}
  \end{figure}
}

Next, we investigate the impact of pilot density on the estimation NMSE. \figref{fig:3bit_ce_diff_alpha} illustrates the NMSE of different estimators as a function of $\alpha=\np/\nt$ when 3-bit ADCs are employed and \SNR{=}{10}. It is observed that the proposed method outperforms all baselines while using less than 50\% pilot overhead. 
We also show the performance of our method when using the orthogonal Zadoff-Chu (ZC) pilot sequence \cite{mo2017channel} with $\alpha=1$, which satisfies the row-orthogonal assumption of $\mathbf{A}$. In contrast, the performance degradation caused by using random QPSK pilots that violate the assumption is minimal, verifying the resilience of our method.

We further show the estimation NMSE with respect to ADC resolutions in \figref{fig:lowbit_ce_diffbit}, where the pilot density is fixed at $\alpha=0.6$. 
The figure reveals that the performance gap between using 1-bit and infinite-bit ADCs 
remains mild at \SNR{\leq}{0}, a desirable region for deploying low-resolution ADCs. Moreover, the proposed method constantly showcases notable gains over the powerful VAE baseline across various ADC resolutions.

\CheckRmv{
  \begin{figure*}[t]
      \centering
      \includegraphics[width=6.5in]{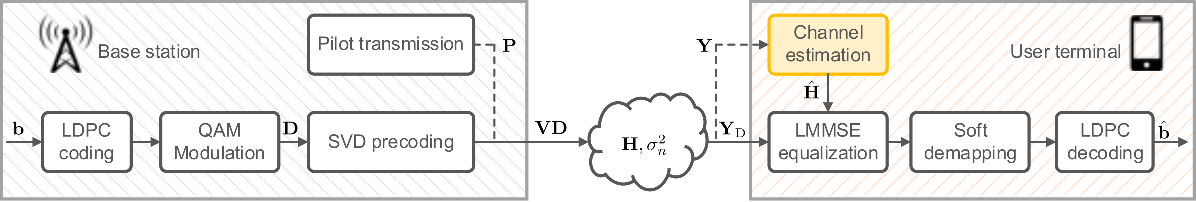}
      \caption{{Block diagram of the considered system setup for evaluating the end-to-end BER performance.}}
      \label{fig:sys}
  \end{figure*}
}

\CheckRmv{
  \begin{figure}[t]
      \centering
      \includegraphics[width=3.2in]{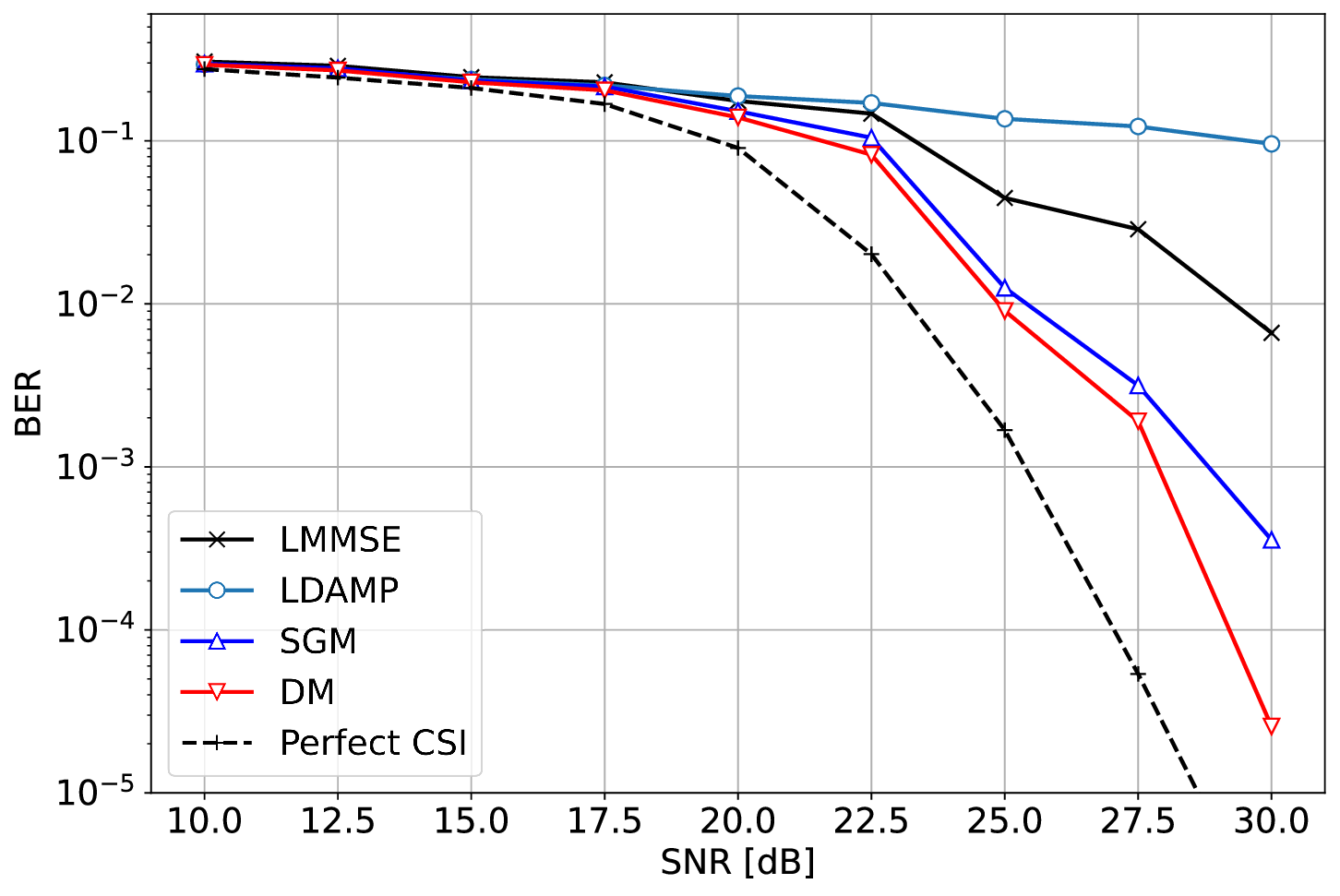}
      \caption{{End-to-end BER performance of the downlink communication link using the proposed estimator.}}
      \label{fig:ber}
  \end{figure}
}

\subsection{{End-to-end Bit Error Rate (BER) Performance}}  \label{sec:ber}

{In this subsection, we evaluate the proposed estimator using the end-to-end BER performance within a downlink transmission system (\figref{fig:sys}) as the metric. The channel setup and the antenna configuration align with the description provided in \secref{sec:simu_setup}. For pilot transmissions, a pilot matrix $\mathbf{P}\in\mathbb{C}^{\nt \times \np}$ with $N_{\rm p}=38$ is sent by the transmitter, and the received signal $\mathbf{Y}$ at the user is utilized for channel estimation, resulting in $\hat{\mathbf{H}}$. During the data transmission phase, a rate-1/2 low-density parity-check (LDPC) code with a codeword size of 648 bits is adopted to encode the information bits $\mathbf{b}$, followed by quadrature amplitude modulation (QAM) to generate the data symbols. The modulation scheme is selected as 64-QAM. The data symbols are partitioned into $N_{\rm s}=8$ streams, denoted as a data matrix $\mathbf{D}$ over multiple channel realizations, and are subsequently precoded for transmission over $N_{\rm t}=64$ antennas. We assume perfect channel state information (CSI) at the transmitter, and the digital precoder $\mathbf{V}$ is constructed by the first $N_{\rm s}$ right-singular vectors of the channel matrix $\mathbf{H}$, obtained through singular value decomposition (SVD).
At the receiver, the data observation $\mathbf{Y}_{\rm D}$ is LMMSE equalized given the estimated channel $\hat{\mathbf{H}}$, soft demapped, and decoded by the LDPC decoder to produce the estimated information bits $\hat{\mathbf{b}}$, with BER computed after $5\times10^4$ codewords are transmitted.}

{\figref{fig:ber} compares the BER performance when different channel estimators are used. The LASSO, EM-GM-AMP, and VAE estimators are not considered in this comparison due to their estimation NMSE floors at high SNRs, which hinders effective data transmission. The end-to-end BER performance with perfect CSI at the receiver is adopted as a performance upper bound for comparison.  
The figure illustrates that our proposed estimator achieves a significant reduction in BER at high SNRs, outperforming the LMMSE- and LDAMP-based approaches by more than 5 dB and surpassing the more complicated SGM-based approach. These findings underscore the significance of the high-fidelity channel acquisition provided by the proposed DM-based estimator in improving end-to-end system performance.}

\subsection{Effectiveness of SURE-DM}  \label{sec:dm_sure}

Previous subsections assume the availability of ground truth channel samples. Next, we target the case where only noisy channel data is accessible to evaluate the effectiveness of the proposed SURE-DM. 
We first train the MMSE denoiser for 100 epochs using the SURE loss given the noisy dataset $\bar{\mathcal{H}}=\{\bar{\mathbf{h}}^{(i)}\}_{i=1}^D$, and then train the DM for 500 epochs using the MMSE-denoised samples corresponding to $\{\bar{\mathbf{h}}^{(i)}\}_{i=1}^D$.

\CheckRmv{
  \begin{figure}[t]
    \centering
    \subfigure[Full-resolution]{
      \includegraphics[width=3.2in]{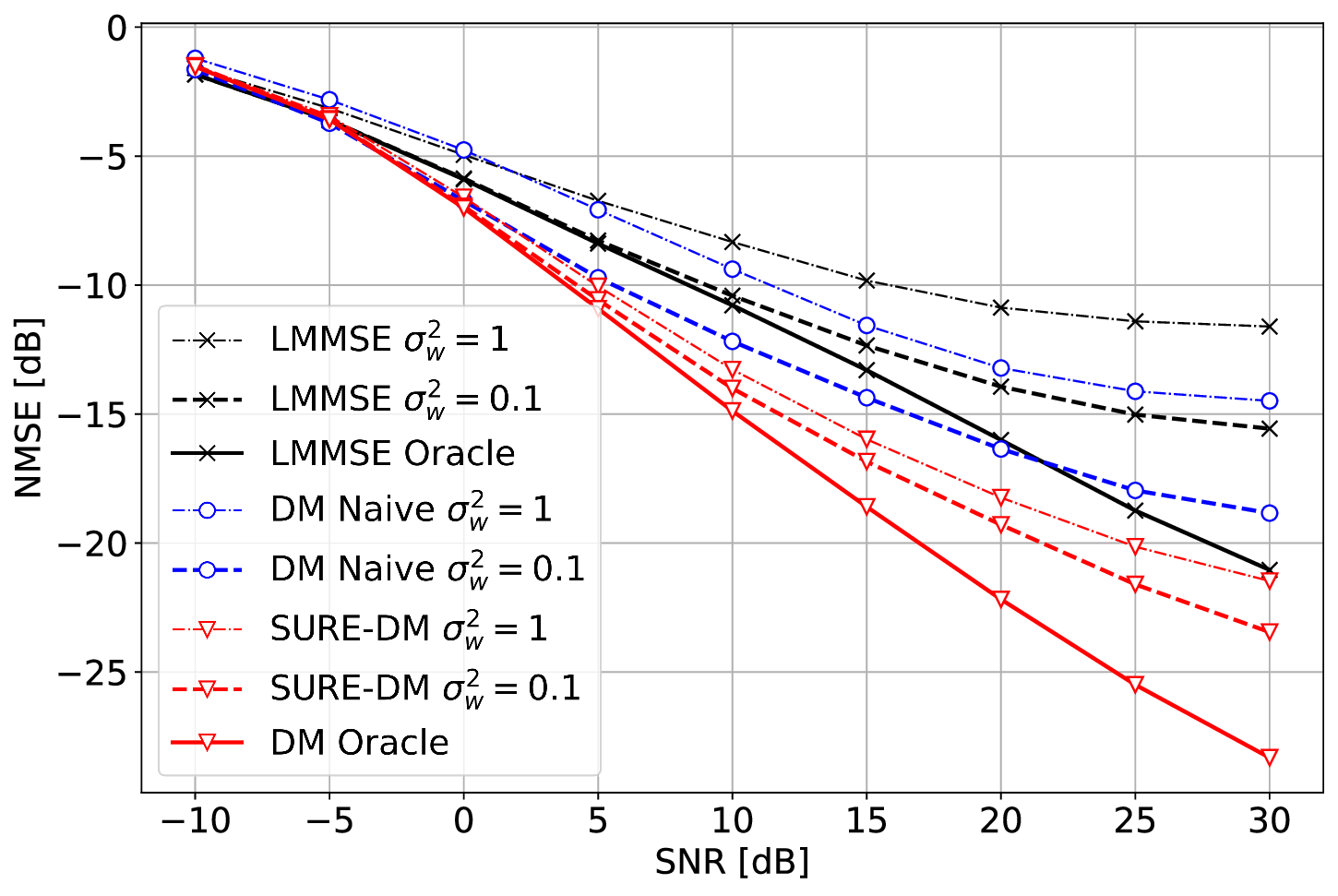}
      \label{fig:noisy_full_ce}
    }
    \subfigure[3-bit quantization]{
      \includegraphics[width=3.2in]{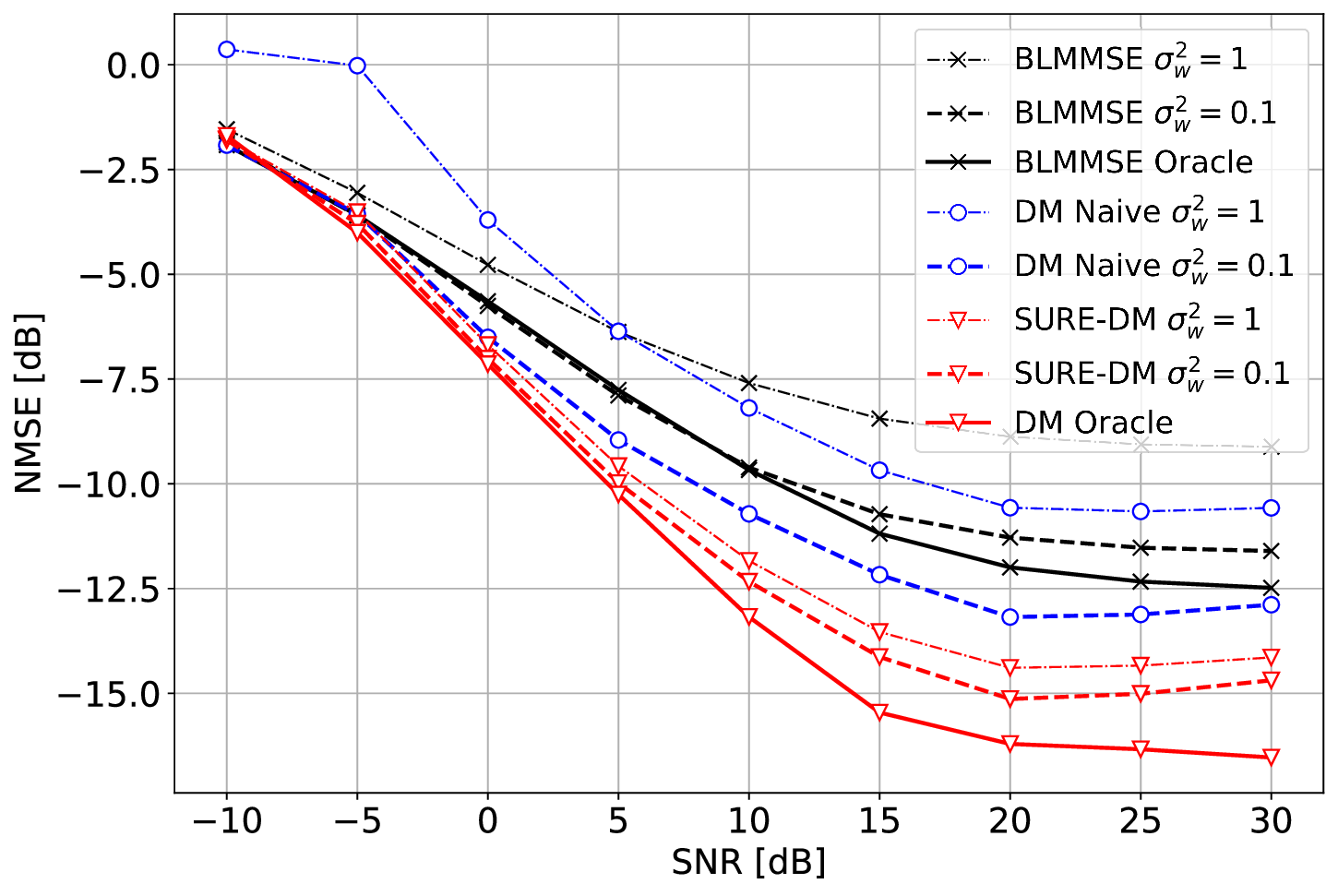}
      \label{fig:noisy_3bit_ce}
    }
    \caption{Performance evaluation of SURE-DM in full-resolution and low-resolution cases.}
    \label{fig:noisy}
  \end{figure}
}

\renewcommand{\arraystretch}{1.2}
\CheckRmv{
  \begin{table*}[t]
    \centering
    \begin{threeparttable}
      \caption{Complexity, Latency, and Parameters Comparison}
    \begin{tabular}{l|c|c|c|c|c|c}
      \hline
      & \multicolumn{2}{c|}{\textbf{FLOPs}} &  \multicolumn{2}{c|}{\textbf{Latency [ms]}}   &        
      \multicolumn{2}{c}{\textbf{Parameters}}                \\
      \hline
      \diagbox{Methods}{$(\nr,\nt)$}        
      & (16, 64)           & (32, 128)               & (16, 64)           & (32, 128)  & (16, 64)               & (32, 128)              \\
      \hline
      BLMMSE \cite{li2017channel}          
      & \textbf{4.9G} &   154.4G             & 14           & 1820 & $1.05\times10^6$ & $1.68\times10^7$ \\
      \hline
      VAE  \cite{fesl2024channel}                  
      & 5.9G  & 133.9G        &  15       &  $>10^4$  & $8.95\times10^6$ & $1.43\times10^8$ \\
      \hline
      SGM  \cite{arvinteMIMOChannelEstimation2022}          
      & 476.8G & 1907.2G               & 280          & 610  &  \multicolumn{2}{c}{$5.89\times10^6$}                  \\
      \hline
      DM              
      & 5.5G  &    \textbf{22.4G}             & \textbf{4.17}            & \textbf{10}   & \multicolumn{2}{c}{$\mathbf{5.50\times10^4}$}  \\
      \hline                                     
    \end{tabular}
    \label{tab:complexity}
    \begin{tablenotes}[para,flushleft]
      \footnotesize
      Note: The FLOPs count and latency is evaluated under 1-bit ADCs and $\alpha=0.6$ and averaged over 100 channel realizations.
    \end{tablenotes}
    \end{threeparttable}
  \end{table*}
}

\figref{fig:noisy} shows the NMSE performance of the model trained using SURE-DM. We adopt noisy training datasets collected at medium and low SNR values, corresponding to $\sigma_w^2\in \{0.1, 1\}$ as illustrated in \secref{sec:sure-dm}. 
The baselines include the LMMSE (BLMMSE) method that computes the channel covariance using noisy training datasets and the DM trained without SURE denoising, marked by ``Naive''. We also consider the LMMSE (BLMMSE) and DM equipped with ground truth channel samples as training datasets, marked by ``Oracle''.

\figref{fig:noisy_full_ce} presents the results in the full-resolution case. LMMSE and the naive DM exhibit notable performance decline, especially when learning at the high noise level of $\sigma_w^2=1$, since they overfit to the noisy data. Our SURE-DM models exhibit markedly enhanced performance by using tailored sequential training, effectively realizing denoising and subsequently learning the structure of the denoised data. Particularly, the SURE-DM with highly noisy data ($\sigma_w^2=1$) even performs on par with the LMMSE using clean data. 
\figref{fig:noisy_3bit_ce} further shows the results when employing 3-bit ADCs, and similar trends as \figref{fig:noisy_full_ce} can be observed.

\subsection{Complexity Analysis} 

We investigate the complexity aspect of the proposed method in this subsection. The metrics we evaluate include the inference floating point operations (FLOPs) count, the latency, and the number of parameters in the model, as shown in \tabref{tab:complexity}.  
We consider two MIMO sizes to analyze the complexity scaling behavior with respect to the antenna count.

The table demonstrates that the FLOPs count of the proposed DM-based estimator is comparable to BLMMSE and VAE at the MIMO size of $(\nr, \nt)=(16, 64)$ and significantly lower than all baselines when $(\nr, \nt)=(32, 128)$.
This result underscores the superior scalability of our method to large-scale MIMO systems due to its linear complexity in proportion to the antenna count. 
For the estimation latency, similar results can be observed: our method achieves a reduction by a factor of 60 at the MIMO size of $(32, 128)$. 
In terms of model parameters, BLMMSE requires $\nr^2\nt^2$ parameters to represent the channel covariance, 
imposing a huge burden as the system dimension increases. 
For SGM and the proposed DM, the input size $(\nr, \nt)$ does not affect the number of parameters due to their fully convolutional nature, a distinct advantage for scaling to large MIMO dimensions. 
Moreover, DM's parameter count is notably lower than SGM due to the lightweight architecture, markedly reducing memory usage.

\vspace{-0.1cm}
\section{Conclusion}
In this paper, we proposed a deep generative prior-aided MIMO channel estimator using denoising diffusion generative models. Based on the prior channel knowledge learned by the DM, a posterior inference method was developed to accurately recover high-dimensional MIMO channels and tackle quantized measurements when low-resolution ADCs are employed. Furthermore, we combined the DM's training with SURE denoising to enable learning from noisy observations. 
Extensive numerical simulations showcase that the proposed method outperforms existing linear, CS-, and DL-based channel estimators in NMSE performance.
It also achieves substantial latency reduction compared to state-of-the-art 
methods in both full- and low-resolution cases. 
This generative learning-based approach demonstrates exceptional scalability and offers a promising solution for high-dimensional channel estimation in next-generation wireless networks.

{Although the proposed method offers significant advantages, further endeavors are needed to address the theoretical analysis on its robustness to out-of-distribution channel data, as well as the fast adaptation under limited data samples. Additionally, theoretical justifications for the heuristic enhancements used in the proposed approach, such as iterative refinement and the gradient scaling parameter, warrant further investigation.}

\appendices
\section{Details of the Denoising Network Structure} \label{appendix:net}
\CheckRmv{
  \begin{figure}[t]
    \centering
    \includegraphics[width=3.45in]{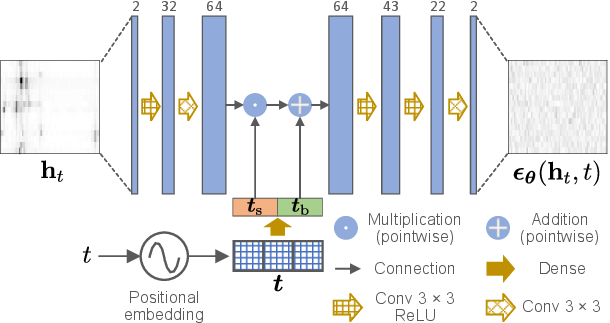}
    \caption{Structure of the DM's denoising network $\boldsymbol{\epsilon}_{\boldsymbol{\theta}}$ using a lightweight CNN with positional embedding of the time step \cite{fesl2024diffusion}. With a little abuse of notations, we use $\mathbf{h}_t$ to denote the matrix form channel sample.}
    \label{fig:network}
  \end{figure}
}

The denoising network (\figref{fig:network}) takes the matrix form of $\mathbf{h}_t$ as the input, treating its real and imaginary parts as two separate convolutional channels.
Aligning with common practices, network parameters are shared across all time steps, and time step index $t$ is specified to the network as a vector $\boldsymbol{t}\in\mathbb{R}^{S_{\rm init}}$ via the Transformer sinusoidal positional embedding \cite{vaswani2017attention}. The time feature vector $\boldsymbol{t}$ is subsequently passed through a fully connected dense layer and the output is divided into the scaling subvector $\boldsymbol{t}_{\rm s}$ and the bias subvector $\boldsymbol{t}_{\rm b}$, both of size $S_{\max}$, the maximum number of convolutional channels in the network. 

Two convolutional layers using kernels of sizes $3\times 3$ initially process the input, with the number of convolutional channels linearly increased to $S_{\max}$, where the first layer employs the rectifier linear unit (ReLU) activation. Next, the time features $\boldsymbol{t}_{\rm s}$ and $\boldsymbol{t}_{\rm b}$ are inserted. Subsequently, three convolutional layers (the first two layers equipped with the ReLU activation) transform intermediate results back to the input size and derive the output via linearly decreasing convolutional channels.

\section{Proof of Proposition~\ref{th:posterior_mean}} \label{appendix:theorem}
We commence with the representation of $p_t(\mathbf{h}_t|\mathbf{y})$ by marginalizing out $\mathbf{h}_{t-1}$ conditioned on $\mathbf{y}$,
\CheckRmv{
  \begin{equation}
    p_t(\mathbf{h}_t|\mathbf{y})=\int p_t(\mathbf{h}_t|\mathbf{h}_{t-1},\mathbf{y})p_{t-1}(\mathbf{h}_{t-1}|\mathbf{y}){\rm d}\mathbf{h}_{t-1}.
    \label{eq:appendix-b1}
  \end{equation}
} 
By taking the gradient with respect to $\mathbf{h}_t$ on both sides, we have
\CheckRmv{
  \begin{align}
    \nabla_{\mathbf{h}_{t}}p_{t}(\mathbf{h}_{t}|\mathbf{y})
    =\int &p_{t-1}(\mathbf{h}_{t-1}|\mathbf{y})\nabla_{\mathbf{h}_{t}}p_{t}(\mathbf{h}_{t}|\mathbf{h}_{t-1},\mathbf{y}){\rm d}\mathbf{h}_{t-1} \nonumber \\
    =\int &p_{t-1}(\mathbf{h}_{t-1}|\mathbf{y})p_{t}(\mathbf{h}_{t}|\mathbf{h}_{t-1},\mathbf{y})\nonumber\\
    & \cdot\nabla_{\mathbf{h}_{t}} \log p_{t}(\mathbf{h}_{t}|\mathbf{h}_{t-1},\mathbf{y}){\rm d}\mathbf{h}_{t-1},
  \end{align}
}
where the last equation originates from the identity $\nabla_x \log f(x) = \nabla_x f(x) / f(x)$. Noting that $\mathbf{h}_{t}$ and $\mathbf{y}$ are conditionally independent given $\mathbf{h}_{t-1}$,  
we replace $\nabla_{\mathbf{h}_{t}} \log p_{t}(\mathbf{h}_{t}|\mathbf{h}_{t-1},\mathbf{y})$ by $\nabla_{\mathbf{h}_{t}} \log p(\mathbf{h}_{t}|\mathbf{h}_{t-1})$ and derive 
\CheckRmv{
  \begin{align}
    \nabla_{\mathbf{h}_{t}}p_{t}(\mathbf{h}_{t}|\mathbf{y}) = \int &p_{t-1}(\mathbf{h}_{t-1}|\mathbf{y})p_{t}(\mathbf{h}_{t}|\mathbf{h}_{t-1},\mathbf{y}) \nonumber \\
    & \cdot \frac{\sqrt{\alpha_t} \mathbf{h}_{t-1} - \mathbf{h}_t}{1-\alpha_t}{\rm d}\mathbf{h}_{t-1}.
    \label{eq:appendix-b2}
  \end{align}
}
where we have utilized the fact that $p(\mathbf{h}_{t}|\mathbf{h}_{t-1})=\mathcal{N}(\mathbf{h}_t;\sqrt{\alpha_t}\mathbf{h}_{t-1},(1-\alpha_t)\mathbf{I})$ according to \eqref{eq:ht-1_to_ht} to obtain the score $\nabla_{\mathbf{h}_{t}} \log p(\mathbf{h}_{t}|\mathbf{h}_{t-1})$. The RHS of \eqref{eq:appendix-b2} can be expanded as 
\CheckRmv{
  \begin{align}
    &\frac{1}{1-\alpha_t}\Big(\int \sqrt{\alpha_t} \mathbf{h}_{t-1} p_{t-1}(\mathbf{h}_{t-1}|\mathbf{y})p_{t}(\mathbf{h}_{t}|\mathbf{h}_{t-1},\mathbf{y}){\rm d}\mathbf{h}_{t-1} \nonumber \\ 
    &- \mathbf{h}_t \int p_{t}(\mathbf{h}_{t}|\mathbf{h}_{t-1},\mathbf{y})p_{t-1}(\mathbf{h}_{t-1}|\mathbf{y}){\rm d}\mathbf{h}_{t-1} \Big) \nonumber \\
    = & \frac{1}{1-\alpha_t}\Big(\sqrt{\alpha_t}p_{t}(\mathbf{h}_{t}|\mathbf{y}) \int \mathbf{h}_{t-1} p_{t-1}(\mathbf{h}_{t-1}|\mathbf{h}_{t},\mathbf{y}){\rm d}\mathbf{h}_{t-1} \nonumber \\
    &- \mathbf{h}_t p_{t}(\mathbf{h}_{t}|\mathbf{y}) \Big)\nonumber \\ 
    = & \frac{p_t(\mathbf{h}_t|\mathbf{y})}{1-\alpha_t} (\sqrt{\alpha_t}\mathbb{E}[\mathbf{h}_{t-1}|\mathbf{h}_{t},\mathbf{y}] - \mathbf{h}_{t}),
    \label{eq:appendix-b3}
  \end{align}
}
where we use the Bayes' rule $p_{t-1}(\mathbf{h}_{t-1}|\mathbf{y})p_{t}(\mathbf{h}_{t}|\mathbf{h}_{t-1},\mathbf{y}) = p_{t-1}(\mathbf{h}_{t-1}|\mathbf{h}_{t},\mathbf{y})p_{t}(\mathbf{h}_{t}|\mathbf{y})$ and \eqref{eq:appendix-b1} to derive the first equality.
Replacing the RHS of \eqref{eq:appendix-b2} with \eqref{eq:appendix-b3}, we have
\CheckRmv{
  \begin{equation}
    \frac{\nabla_{\mathbf{h}_{t}}p_{t}(\mathbf{h}_{t}|\mathbf{y})}{p_{t}(\mathbf{h}_{t}|\mathbf{y})} = \frac{1}{1-\alpha_t} (\sqrt{\alpha_t}\mathbb{E}[\mathbf{h}_{t-1}|\mathbf{h}_{t},\mathbf{y}] - \mathbf{h}_{t}).
  \end{equation}
}
Recalling the identity $\nabla_x \log f(x) = \nabla_x f(x) / f(x)$, we get
\CheckRmv{
  \begin{equation}
    \nabla_{\mathbf{h}_{t}}\log p_{t}(\mathbf{h}_{t}|\mathbf{y}) = \frac{1}{1-\alpha_t} (\sqrt{\alpha_t}\mathbb{E}[\mathbf{h}_{t-1}|\mathbf{h}_{t},\mathbf{y}] - \mathbf{h}_{t}).
  \end{equation}
} 
The conditional posterior mean $\mathbb{E}[\mathbf{h}_{t-1}|\mathbf{h}_{t},\mathbf{y}]$ as given in the RHS of \eqref{eq:condition_post} can be derived by rearranging this result, thereby completing the proof.

\section{Proof of Proposition~\ref{quantized_likelihood}} \label{appendix:quan}  
Under the assumption that $\mathbf{A}$ is row-orthogonal such that $\mathbf{AA}^{T}$ is a diagonal matrix, each element $\tilde{n}_m$ of the effective noise $\tilde{\mathbf{n}}$ independently follows $\mathcal{N}\left(0,\tilde{\sigma}_m^2\right)$, where $\tilde{\sigma}_m^2 \triangleq \frac{1-\bar{\alpha}_t}{\bar{\alpha}_t}\|\mathbf{a}_m^T\|_2^2+\sigma_n^2$. Hence, \eqref{eq:pseudo_likelihood_quan} can be simplified as 
\CheckRmv{
  \begin{equation}
    p_t(\bar{\mathbf{y}}|\mathbf{h}_t)=\prod_{m=1}^M p_{\tilde{n}}\big(z_m+\tilde{n}_m \in Q^{-1}(\bar{y}_m)\big),
    \label{eq:appendix-c1}
  \end{equation}
}
where $p_{\tilde{n}}(\cdot)$ is the probability density function of $\tilde{n}$, with the index $m$ omitted for brevity. Based on the definition of the quantizer $Q$, we have 
\CheckRmv{
  \begin{align}
    p_{\tilde{n}}(z_m+\tilde{n}_m&\in Q^{-1}(\bar{y}_m)) \nonumber\\
    = & {p_{\tilde{n}}(\bar{y}_{m}^\mathrm{low}\leq z_m+\tilde{n}_m<\bar{y}_{m}^\mathrm{up})}\triangleq P_m.
    \label{eq:appendix-c2}
  \end{align}
}
Since $\tilde{n}_m\sim \mathcal{N}(0,\tilde{\sigma}_m^2)$, it is straightforward that
\CheckRmv{
  \begin{equation}
    P_{m} 
    = \Phi(\tilde{y}_{m}^{\rm {up}}) -\Phi(\tilde{y}_{m}^{\rm {low}}),
    \label{eq:appendix-c3}
  \end{equation}
}
where $\tilde{y}_{m}^{\rm {up}}$ and $\tilde{y}_{m}^{\rm {up}}$ are given in \eqref{eq:modified_threshold}.

Based on \eqref{eq:appendix-c1}-\eqref{eq:appendix-c3}, the score of $p_t(\bar{\mathbf{y}}|\mathbf{h}_t)$ is computed as
\CheckRmv{
  \begin{align}
    \nabla_{\mathbf{h}_t} &\log p_t(\bar{\mathbf{y}}|\mathbf{h}_t) \nonumber\\
    =& \sum_{m=1}^M \nabla_{\mathbf{h}_t}z_m \cdot \nabla_{z_m} \log p_{\tilde{n}}\big(z_m+\tilde{n}_m \in Q^{-1}(\bar{y}_m)\big) \nonumber \\
    =& \sum_{m=1}^M \frac{1}{\sqrt{\bar{\alpha}_t}} \mathbf{a}_m \underbrace{\frac{\partial \log P_{m}}{\partial z_m}}_{g_{m}}, 
    \label{eq:appendix-c4}
  \end{align}
}
where we have utilized the fact that $z_m=\frac{\mathbf{a}_m^T\mathbf{h}_t}{\sqrt{\bar{\alpha}_t}}$ in the last equality.
By taking partial derivatives of \eqref{eq:appendix-c3} with respect to $z_m$, the expression for $g_{m}$ given in \eqref{eq:g_real} can be verified. Therefore, by writing \eqref{eq:appendix-c4} into a compact form, we derive the score given in \eqref{eq:pseudo_likelihood_quan_score}, completing the proof.



\ifCLASSOPTIONcaptionsoff
  \newpage
\fi




\end{document}